\documentclass{pazha2}
\usepackage{graphicx}
\usepackage{multirow}
\usepackage{pdflscape} 
\usepackage{amsmath}
\usepackage{color}
\definecolor{darkblue}{rgb}{0,0,0.9}

\def\*{$^{*}$}
\def\aa{$^{\mbox{\small a}}$}
\def\bb{$^{\mbox{\small b}}$}
\def\cc{$^{\mbox{\small c}}$}
\def\dd{$^{\mbox{\small d}}$}
\def\ee{$^{\mbox{\small e}}$}
\def\ff{$^{\mbox{\small f}}$}
\def\g2{$^{\mbox{\small g}}$}

\begin{document}
\journalinfo{2019}{45}{12}{791}{835}{865}[820]
\sloppypar
%\baselineskip 21pt

%\title{\bf Decrease in the Brightness of the Cosmic X-ray and
%  Soft Gamma-ray Background toward Clusters of Galaxies}
\title{\bf DECREASE IN THE BRIGHTNESS OF THE COSMIC X-RAY AND SOFT GAMMA-RAY
BACKGROUND TOWARD CLUSTERS OF GALAXIES}
\year=2019
\author{
S. A.~Grebenev\address{1}\email{grebenev@iki.rssi.ru} and
R. A.~Sunyaev\address{1,2}
\addresstext{1}{Space Research
    Institute, Russian Academy of Sciences, Profsoyuznaya ul. 
    84/32, Moscow, 117997 Russia}
\addresstext{2}{Max-Planck-Institut f\"{u}r Astrophysik,
  Karl-Schwarzschild-Str.\,1, Postfach\,1317, D-85741 Garching,\,Germany}
}

\shortauthor{GREBENEV, SUNYAEV}
\shorttitle{DECREASE IN THE BRIGHTNESS OF THE COSMIC X- AND
  GAMMA-RAY BACKGROUND} 

\submitted{February 25, 2019}
\revised{September 18, 2019}
\accepted{October 23, 2019}
%\published{\today}

%\vspace{2mm}
\begin{abstract}
\noindent
We show that Compton scattering by electrons of the hot
intergalactic gas in galaxy clusters should lead to peculiar
distortions of the cosmic background X-ray and soft gamma-ray
radiation --- an increase in its brightness at $h\nu\la
60$--$100$ keV and a drop at higher energies. The background
distortions are proportional to the cluster gas surface density,
in contrast to the intensity of the thermal gas radiation
proportional to the density squared, which allows the most
important cluster parameters to be measured. The spectral shape
of the background distortions and its dependence on the gas
temperature, optical depth, and surface density distribution law
have been studied using detailed Monte Carlo computations and
confirmed by analytical estimations. In the cluster frame the
maximum of the background decrease due to the recoil effect
occurs at $h\nu\sim500$--$600$ keV. The photoionization of
hydrogen- and helium-like iron and nickel ions leads to
additional distortions in the background spectrum --- a strong
absorption line with the threshold at $h\nu\sim9$ keV (and also
to an absorption jump at $h\nu\ga2$ keV for cold clusters). The
absorption of intrinsic thermal radiation from the cluster gas
by these ions also leads to such lines. In nearby ($z\la1$)
clusters the line at $h\nu\ga2$ keV is noticeably enhanced by
absorption in the colder ($\sim10^6$ K) plasma of their
peripheral ($\la 3$ Mpc) regions; moreover, the absorption line
at $h\nu\sim 1.3$ keV, which does not depend on the properties
of the hot cluster gas, splits off from it. The redshift of
distant clusters shifts the absorption lines in the background
spectrum (at $\sim2$, $\sim9$, and $\sim500$ keV) to lower
energies. Thus, in contrast to the microwave background
radiation scattering effect, this effect depends on the cluster
redshift $z$, but in a very peculiar way. When observing
clusters at $z\ga1$, the effect allows one to determine how the
X-ray background evolved and how it was ``gathered'' with
$z$. To detect the effect, the accuracy of measurements should
reach $\sim0.1$\%. We consider the most promising clusters for
observing the effect and discuss the techniques whereby the
influence of the thermal gas radiation hindering the detection
of background distortions should be minimal.\\

\noindent
{\bf DOI:} 10.1134/S1063773719120016\\    

{\bf Keywords:\/} {cosmic background radiation, galaxy clusters, hot and
  warm-hot intergalactic plasma, Compton scattering, recoil
  effect, Doppler effect, photoionization, bremsstrahlung and
  recombination radiation.}
\end{abstract}

%xxxxxxxxxxxxxxxxxxxxxxxxxxxxxxxxxxxxxxxxxxxxxxxxxxxxxxxxxxxxxxx
\section{INTRODUCTION}
\noindent
In recent years the effect of a decrease in the brightness of
the cosmic microwave background radiation toward galaxy clusters
has turned from an elegant theoretical idea (Sunyaev and
Zel'dovich 1970, 1972, 1980, 1981; Sunyaev 1980; Zel'dovich and
Sunyaev 1982) into one of the most important tools for studies
in the field of observational cosmology and astrophysics of the
early Universe. The energy redistribution of photons in the
background radiation spectrum after their Thomson scattering by
electrons of the hot ($kT_{\rm e}\sim 2-15$ keV) intergalactic
cluster gas underlies this effect. In this case, a deficit of
photons is formed in the low-frequency part of the spectrum (at
energies $h\nu\la3.83\ kT_{\rm r},$ where $T_{\rm r}\simeq 2.7$
K is the current temperature of the cosmic microwave background
radiation), i.e., a ``negative'' source (a ``hole'' in the
background) appears, while a bright ``positive'' source with an
unusual spectrum appears in the high frequency part. This effect
is unique in that its action is determined by the optical depth
of the cluster gas for scattering by electrons along the line of
sight $\tau_{\rm T}=\sigma_{\rm T}\int N_{\rm e}(l)\,{\rm
  d}\,l$, i.e., it is proportional to the gas density and not to
the density squared, as the brightness of the intrinsic thermal
radiation from the hot gas. Here, $\sigma_{\rm T}$ is the
Thomson scattering cross section. Surprisingly, the amplitude of
the effect does not decrease with cluster distance (redshift
$z$); the spectral shape of the background distortions does not
depend on $z$ either. Owing to these properties, the effect is
widely used to determine the parameters of clusters and to
effectively search for them. The effect is successfully observed
with the specially constructed SPT (South Pole Telescope,
Carlstrom et al. 2002; Williamson et al. 2011; Bleem et
al. 2015) and ACT (Atacama Cosmology Telescope, Hasselfield et
al. 2013) telescopes and a number of other telescopes
(Birkinshaw 1999); the Planck satellite (Ade et al. (Planck
Collaboration) 2014, 2015, 2016a) has made an enormous
contribution to the investigation of the effect. 

In this paper we consider a similar effect --- the distortions
arising due to scattering by electrons of the hot gas of
clusters in the cosmic X-ray background.  The existence of such
an effect has already been mentioned in Sunyaev and Zel'dovich
(1981), Zel'dovich and Sunyaev (1982), and Khatri and Sunyaev
(2019).  Based on simple nonrelativistic ($h\nu \ll m_{\rm
  e}c^2$) estimates, these authors concluded that it is
impossible to directly observe the effect against the intrinsic
thermal X-ray background of the cluster gas. At the same time,
they noted that it is potentially important to take into account
this effect when considering the thermal balance of the gas. In
this paper we performed relativistically accurate Monte Carlo
computations of the effect and investigated and discussed the
prospects for its observation in a harder ($h\nu \ga 60$ keV)
energy range.

The X-ray background (diffuse) radiation differs from the cosmic
microwave background radiation by its origin (it is a
superposition of the radiation spectra of a large number of AGNs
--- active galactic nuclei and quasars), but, at the same time,
it is also characterized by a high degree of isotropy and
homogeneity.  In the standard ($h\nu\la10$ keV) X-ray range from
60 to 80\% of the background radiation has already been resolved
into separate (point) sources by telescopes with
grazing-incidence mirrors (see, e.g., Hasinger et al. 1998;
Miyaji et al. 2000; Giacconi et al.  2001); another $\sim10$\%
of the background has been explained by the thermal radiation of
the gas in galaxy clusters. An extrapolation of the X-ray
spectra for detected AGNs to the harder ($h\nu\ga10$ keV) energy
range, which takes into account the evolution of their number
and the composition of their population with $z$, shows that
here they should also make a dominant contribution to the
background spectrum (Sazonov et al. 2008; Ueda et al. 2014;
Miyaji et al. 2015). It follows from the simulations by Ueda et
al. (2014) that the observed background spectrum is formed
mainly at $z\sim1$ or slightly farther, but its shape changes
little even at higher $z$, while the intensity (in the rest
frame) decreases slowly (in the range $z \sim 1.5-3$ the
emissivity of AGNs remains almost constant, see Miyaji et
al. 2000; Ueda et al. 2014). To a first approximation, up to
$z\la2$ we may neglect these changes and assume that the X-ray
background, along with the cosmic microwave one, is subject only
to the ordinary cosmological expansion (see, e.g., Madau and
Efstathiou 1999). The increase in the fraction of AGNs with high
luminosities at $z\ga1$ is of greater importance. This should
lead to an enhancement of the background fluctuations. Later we
will discuss whether the appearance of such AGNs affects the
detection of the effect under consideration.

The action of Compton scattering on cosmic microwave and X-ray
background photons also differs: when the cosmic microwave
background radiation is scattered, the electron temperature
exceeds the photon energy and, therefore, the photons gain
energy due to the Doppler effect; when the X-ray radiation is
scattered, the photon energy exceeds (or at least is comparable
to) the electron temperature and, accordingly, the photons, as
will be shown below, on average, lose their energy due to the
recoil effect.

%xxxxxxxxxxxxxxxxxxxxxxxxxxxxxxxxxxxxxxxxxxxxxxxxxxxxxxxxxxxxxxx
\section*{THE X-RAY BACKGROUND SPECTRUM}
\begin{figure*}[t]
\begin{minipage}{0.50\textwidth}
  \includegraphics[width=1.0\textwidth]{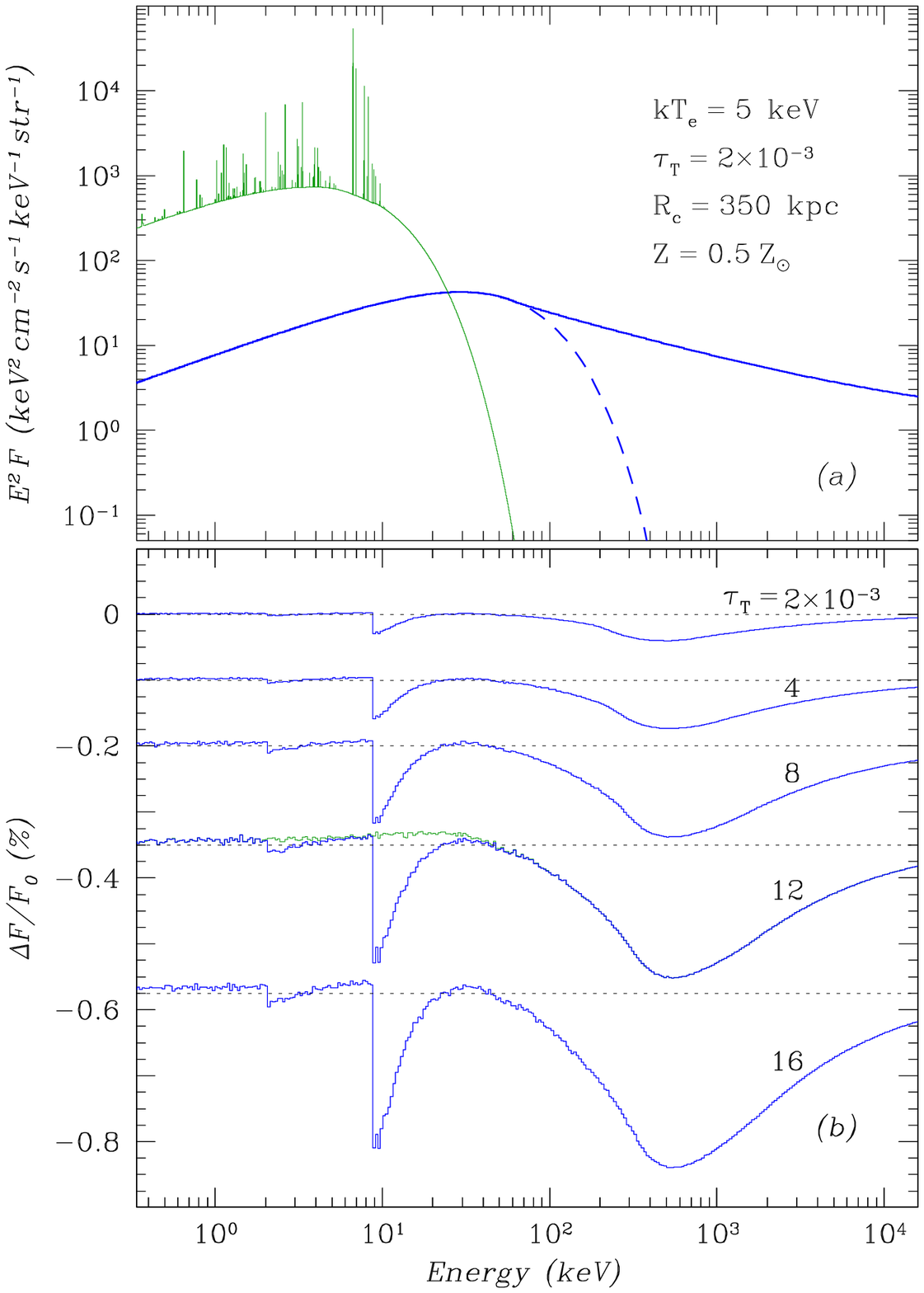}
%\epsffile{szxray_tau.ps}
\end{minipage}\begin{minipage}{0.50\textwidth}
 \includegraphics[width=1.0\textwidth]{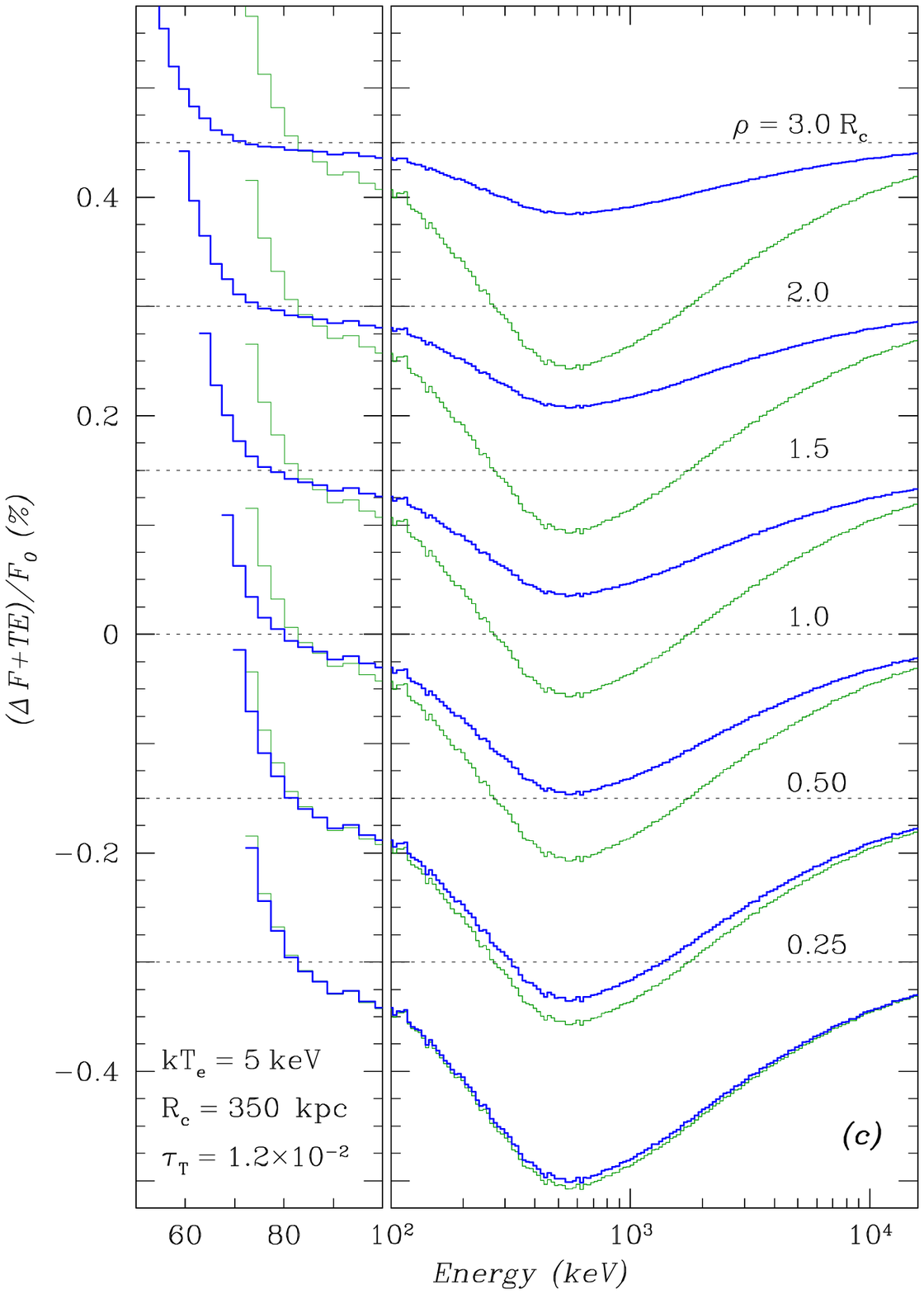}
%\epsffile{szxray_ang2.ps}
\end{minipage}

\caption{\rm (a) Spectra of the cosmic X-ray background (solid
  thick blue line) and the thermal (bremsstrahlung and
  recombination) X-ray radiation from the hot intergalactic gas
  (thin green line, 5-eV resolution) of a cluster with a core
  radius $R_{\rm c}=350$ kpc and uniform electron temperature
  and density distributions inside it, $kT_{\rm e}=5$ keV and
  $N_{\rm e}=1.4\times10^{-3}\ \mbox{cm}^{-3}$ (so that
  $\tau_{\rm T}=2\times10^{-3}$), the gas metallicity is
  $Z=0.5\,Z_{\odot}$. The dashed line indicates the exponential
  extension of the background fit at low ($h\nu\la 60$ keV)
  energies. (b) The relative background distortions due to
  scattering by cluster gas electrons with the same temperature
  $kT_{\rm e}=5$ keV, but different optical depths for Thomson
  scattering, $\tau_{\rm T}=2,\ 4,\ 8,\ 12\ \mbox{and}\ 16\times
  10^{-3}$ (along the line of sight passing through the cluster
  center). (c) Attenuation of the distortions when they are
  observed at different impact parameters
  $\rho=0.25,\ 0.5,\ 1.0,\ 1.5,\ 2.0\ \mbox{and}\ 3.0\ R_{\rm
    c}$ relative to the center for a cluster with a $\beta$
  density distribution with $N_{\rm c}=7.5\times
  10^{-3}\ \mbox{cm}^{-3}$ and the same core radius and
  temperature as those on other panels. The thin (green) lines
  indicate the distortion profile at the center. The thermal gas
  radiation is included in the distortions. The dotted straight
  lines correspond to the unperturbed background for each
  case.\label{fig:tau}}
\vspace{-2mm}
\end{figure*}

\noindent
The broadband spectrum of the hard X-ray ($h\nu>3$ keV) cosmic
background radiation was measured with the instruments of the
HEAO-1 observatory (Gruber et al. 1999a). Although the absolute
normalization of the spectrum was subsequently the subject of
discussion (the INTEGRAL observations gave a value greater by
$\sim10$\%, Churazov et al. 2007), the shape of the background
spectrum was confirmed.  We will use the following fit to the
background spectrum (Gruber et al. 1999a):
$$\setcounter{equation}{1}
S_0(E)\simeq \left \{ 
%\begin{array}{ll}
%0.55\ e^{x/2}\ \left[\, \ln(6.1/x)\,
%(1+x^2/16)-0.98\,\right],& \mbox{если} \ x < 1.7\,\\
%0.55(x/2)^{-0.5}\left[1.25-(0.312-0.24/x)/x\right] \mbox{~~~},& \mbox{если} \ x
%\geq 1.7.\\ 
%\end{array}\right.
\begin{array}{ll}
7.877\ E^{-0.29} e^{-{E}/{41.13}},
  & \mbox{at} \ E < 60\\
  0.0259\ (E/60)^{-5.5}+& \makebox[22mm]{}(1)\\ 
0.504\ (E/60)^{-1.58}+ \mbox{~~},&\mbox{at} \ E > 60.\\
  0.0288\ (E/60)^{-1.05}\\
\end{array}\right.
$$ Here, $S_0(E)$ is the energy flux expressed in keV $\mbox{cm}^{-2}\ \mbox{s}^{-1}\ \mbox{keV}^{-1}\ \mbox{sr}^{-1}$ and
$E$ is the photon energy $h\nu$ in keV. This fit agrees well
with the gamma-ray background measurements at energies 1 MeV --
100 GeV by the COMPTEL and EGRET telescopes of the CGRO
observatory. The undistorted background spectrum corresponding
to this fit is indicated in Fig.\,\ref{fig:tau}a by the thick
solid (blue) line. The dotted line indicates the spectrum
corresponding to the extension of the soft component in Eq. (1)
to the hard range.

As we will see below, the distortions in this spectrum that
arise when the background radiation interacts with electrons of
the hot gas in a galaxy cluster are fairly small and do not
exceed the current accuracy of our knowledge of the spectrum
shape and parameters. In this regard the effect being discussed
seems more difficult to measure than the distortion of the
microwave background radiation whose spectrum has an almost
ideal Planckian shape. However, if the accuracy of X-ray and
gamma-ray measurements will increase to the required level in
future, then the undistorted spectrum will be remeasured
simultaneously with the distorted one, which will allow the
deviations to be revealed. The estimates of the relative
deviations of the background spectrum presented in this paper
remain valid.
%---------------------------------------------------------------------------------------------
%---------------------------------------------------------------------------------------------

%xxxxxxxxxxxxxxxxxxxxxxxxxxxxxxxxxxxxxxxxxxxxxxxxxxxxxxxxxxxxxxx
\section*{MONTE CARLO COMPUTATIONS}
\noindent
As the initial approximation we will assume the hot gas in a
cluster to be distributed spherically symmetrically with uniform
electron density $N_{\rm e}$ and temperature $kT_{\rm e}$ within
its radius $R_{\rm c}$. The optical depth of such a gas cloud
for scattering calculated along the line of sight passing
through its center will be $\tau_{\rm T}=2 \sigma_{\rm T} N_{\rm
  e} R_{\rm c} = 2\tau_{\rm c}$. Our computations of the Compton
scattering of the background radiation in such a cloud were
performed by the Monte Carlo method in accordance with the
algorithms developed by Pozdnyakov et al. (1983). The background
radiation was assumed to be incident on the cloud
isotropically. The angle-averaged radiation leaving the cloud
was considered as the emergent one. In this sense,
characterizing below the amplitude of the distortions in the
spectrum by the optical depth along the line of sight passing
through the cloud center, by $\tau_{\rm T}$ we mean the
characteristic of the cloud itself. The gas cloud with an
optical depth $\tau_{\rm T}=1\times10^{-2}$ has a mass $M_{\rm
  g}\simeq 3.1\times10^{13}\ (\tau_{\rm T}/0.01)(r_{\rm
  c}/350\ \mbox{\rm kpc})^2 M_{\odot}.$ The total mass $M_{500}$
of the corresponding cluster, including the dark matter, should
be greater at least by an order of magnitude. This is a
moderate-mass cluster like the Coma cluster. Below we will also
consider more massive clusters (see Table\,\ref{table:clusters}).

The hydrogen and helium in the cluster gas were assumed to have
normal cosmic abundances, $X\simeq 0.74$ and $Y\simeq 0.24$ by
mass (Allen et al. 1973), respectively, $N_{\rm
  e}\simeq(X+0.5\,Y)\,\rho/m_{\rm p}\simeq 0.86\,\rho/m_{\rm
  p}$. As a rule, the abundance of the iron-group elements was
taken to be $Z=0.5\,Z_{\odot},$ but it could change. At a gas
temperature typical for clusters the atoms of most elements are
ionized fully, iron is ionized to the hydrogen- and helium-like
states, and nickel is ionized to the lithium-like
state. Photoabsorption by Fe\,XXVI and Fe\,XXV ions introduces
distortions into the spectrum comparable in relative amplitude
to the distortions due to scattering by electrons. Therefore,
this process should be taken into account in the computations.

The degree of iron ionization for a plasma of the required
temperature was obtained from the code by Raymond and Smith
(1977), which is used to compute the ionization balance of an
optically thin plasma. Apart from the Fe\,XXV and Fe\,XXVI ions,
the absorption by Fe\,XXII -- Fe\,XXIV ions was taken into
account when computing the distortions of the background
spectrum. The absorption by Ni\,XXIII -- Ni\,XXVIII\footnote{In
  our computation of the photoabsorption in a warm-hot
  intergalactic medium with $kT_{\rm e}=0.2$ keV we took into
  account the Fe\,XIV -- Fe\,XVII and Ni\,XIII -- Ni\,XVII ions.} was
taken into account in the same way. The ionization by background
photons in the ionization balance was neglected. We used the
fits of the cross sections for photoabsorption by various ions
from Verner and Yakovlev (1995) and Verner et al. (1996). The
bremsstrahlung and recombination radiation spectrum of the
intergalactic cluster plasma was also computed using the
Raymond-Smith code.  The thin solid (green) line in
Fig.\,\ref{fig:tau}a indicates this spectrum for a cluster with
a plasma temperature $kT_{\rm e}=5$ keV. On the whole, the
recombination onto the iron atoms ionized as a result of cosmic
background photoabsorption is analogous to the recombination
onto the ions formed by collisional processes.  The
corresponding recombination radiation gives only a small
$\la1$\% contribution to the intensity of the recombination
radiation predicted by the Raymond-Smith code and does not
change its spectrum.

%===========================================
\subsection{The Model with a Uniform Density Distribution}
\noindent
Figure\,\ref{fig:tau}b presents the distortions in the
background radiation spectrum (Eq. (1)) computed for various
optical depths of the cloud\footnote{Or various electron
  densities $N_{\rm e}$, because at a fixed cluster core radius
  the optical depth and density are related uniquely, $\tau_{\rm
    T}\sim N_{\rm e}$.}. The plasma temperature was assumed for
all computations to be the same and equal to $kT_{\rm e}=5$ keV,
the heavy element abundance was $Z=0.5\,Z_{\odot}$. It can be
seen from the spectrum corresponding to $\tau_{\rm T}=1.2\times
10^{-2}$, for which the thin (green) line indicates the result
of our computation without absorption by iron ions, that Compton
scattering slightly (by $\sim 0.02$\%) raises the background
intensity at energies $h\nu\la 40$ keV.  However,
photoabsorption reduces this rise beyond the iron ionization
threshold $h\nu\ga 9$ keV (blue line).  In contrast, a dramatic
drop in intensity reaching $\sim0.2$\% near the threshold is
observed in the background spectrum due to absorption at these
energies.  The maximal increase and decrease in intensity in the
spectrum are $\la0.005$ and $\sim0.03-0.04$\% for the spectrum
corresponding to the smallest optical depth $\tau_{\rm
  T}=2\times 10^{-3}$ of those considered and reach $\sim0.03$
and $\sim0.23$\%, respectively, for the largest optical depth
$\tau_{\rm T}=1.6\times 10^{-2}$. The background spectrum also
exhibits a weak absorption line on the $L$ shell of iron ions
with the threshold at $h\nu\ga 2$ keV and an amplitude of
$\sim0.04$\% (for a cluster with $\tau_{\rm T}=1.6\times
10^{-2}$).
%---------------------------------------------------------------------------------------
\begin{table}[tb]
  \caption{Parameters of the MeV dip in the background
spectrum toward a cluster with a uniform gas
density\aa\label{table:edge.mev}} 
  \vspace{-2mm}
 
\begin{center} \small
  \begin{tabular}{c|c|c|c|c}\hline\hline
    $\tau_{\rm T},$& \ \ \ $h\nu_{\rm th}$\bb \ &
    \ \ \ $h\nu_{\gamma}$\cc \ &
    \ \ $h\Delta\nu_{\gamma}$\dd&$W_{\gamma}$\ee\\ \hline
    & & & & \\ [-2mm]
   $\times 10^{-3}$& keV&  keV& MeV&keV\\ \hline
            & & & & \\ [-2mm]
    $2$  &29&584&2.3&  1.1\\%225.642410-2531.74927
    $4$  &33&584&2.7&  3.4\\%203.432693-2906.83740
    $8$  &35&584&2.9&10.0\\%189.854431-3114.73193 
    $12$&33&584&3.0&18.1\\%189.854431-3224.19067
    $16$&35&604&3.0&27.1\\%196.526321-3224.19067
    \hline
\multicolumn{5}{l}{}\\ [-1mm]
\multicolumn{5}{l}{\aa\ $kT_{\rm e}=5$ keV.}\\
\multicolumn{5}{l}{\bb\ The energy of the beginning of the dip.}\\  
\multicolumn{5}{l}{\cc\ The energy of the deepest point of the dip.}\\  
\multicolumn{5}{l}{\dd\ The FWHM of the dip.}\\  
\multicolumn{5}{l}{\ee\ The equivalent width of the dip.}
\end{tabular}
\end{center}
\vspace{-6mm} 
\end{table}
%---------------------------------------------------------------------------------------

At energies $h\nu\ga 60$ keV scattering by electrons leads to a
``dip'' in the background spectrum due to the recoil effect. As
a result of this effect, the photons lose a certain fraction of
their energy and are shifted downward along the frequency
axis. These shifted photons make a certain contribution to the
intensity excess in the spectrum at energies $h\nu\la 60$ keV
(to be more precise, at $h\nu\la kT_{\rm e}$, see the
low-frequency asymptotics to Eq. (\ref{eq:relintpl}) below and
the curve corresponding to $kT_{\rm e}=0$ keV in
Fig.\,\ref{fig:komp}). The soft photons shifted upward along the
frequency axis (to $h\nu\sim 3 kT_{\rm e}$) due to the Doppler
effect make a major contribution to the excess.
%---------------------------------------------------------------------------------------------
\begin{figure}[t]
  \vspace{-1mm} 
\centerline{\includegraphics[width=1.05\linewidth]{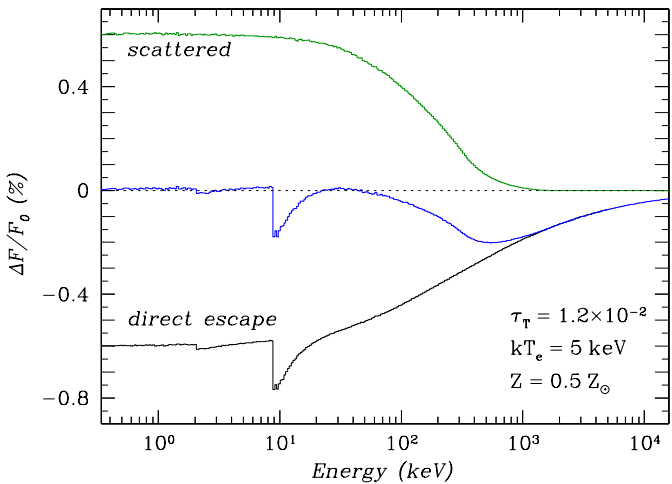}}

%\epsfxsize=1.05\linewidth
%\epsffile{szxray_single.ps}
%\fbox{\rule{0cm}{6cm}\rule{0.97\linewidth}{0cm}}
\caption{\rm {\color{black} Comparison of the background
    distortions arising in the hot cluster gas (relative to its
    initial spectrum) in the direct escape radiation (black
    curve) and scattered radiation (green curve). The blue
    (thick) curve indicates the sum of these distortions. The
    cluster gas is assumed to be distributed uniformly within
    the radius $R_{\rm c}=350$ kpc and to have an optical depth
    $\tau_{\rm T}=1.2\times10^{-2}$, temperature $kT_{\rm e}=5$
    keV, and metallicity $Z=0.5\,Z_{\odot}$}. \label{fig:1scat}}
\vspace{-2.mm} 
\end{figure}
%---------------------------------------------------------------------------------------------

In the range $h\nu\sim 500-600$ keV, where the depth of the dip
in the background spectrum due to the recoil effect is maximal,
the drop in background radiation brightness reaches $0.1-0.2$\%
for realistic optical depths of the cluster, $\tau_{\rm T}\sim
(4-12)\times 10^{-3}.$ At $\tau_{\rm T}=1.6\times 10^{-2}$ the
dip deepens to $\sim0.25$\%.

In Table\,\ref{table:edge.mev} the depth of the MeV dip is given
in terms of the equivalent width $W_{\gamma}$ for various
Thomson optical depths $\tau_{\rm T}$ of the gas. Given that the
line is broad and the background intensity along its profile can
change greatly, we defined the equivalent width as follows:
$W_{\gamma}=\int^\infty_{\nu_{\rm
    th}}\,(F_0-F_{\nu})/F_0\,d\,h\nu$, where $F_0(\nu)$ is the
photon spectrum of the background
[$S_0(\nu)=h\nu\,F_0(\nu)$]. We considered the same model
cluster with a uniform density, $R_{\rm c}=350$ kpc, and
$kT_{\rm e}=5$ keV. The table also gives the energies of the
beginning, $h\nu_{\rm th},$ and the center, $h\nu_{\gamma},$ of
the dip in the spectrum and its full width at half maximum
(FWHM) $h\Delta\nu_{\gamma}$. Whereas the $h\nu_{\rm th}$ and
$h\nu_{\gamma}$ variations in Table\,\ref{table:edge.mev} can be
attributed to the error of our calculation, the changes in
$h\Delta\nu_{\gamma}$ and $W_{\gamma}$ are real, they reflect a
drop in the dip amplitude as $\tau_{\rm T}$ decreases.

Note that the dips in the background spectrum
related to the recoil effect and photoabsorption are
formed in the radiation going through the cluster from
its back side. To a first approximation, their depth is
proportional to the mean optical depth of the cluster
gas in this direction:
\begin{equation}\label{eq:untauav}
<\!\tau_{\rm T}\!> = \frac{2}{R_{\rm c}^2} \int^{R_{\rm c}}_0
\tau_{\rm T}(\rho) \rho\,d\rho= \frac{2}{3}\,\tau_{\rm T},
\end{equation}
where $\tau_{\rm T}(\rho)$ is the Thomson optical depth of the
cluster along the line of sight passing at an impact
parameter $\rho$ from the center. It is 
\begin{equation}\label{eq:untau}
  \tau_{\rm T}(\rho)=\left \{
  \begin{array}{ll}
    \tau_{\rm  T} \left(1-\rho^2/\,R_{\rm
      c}^2\right)^{1/2},& \mbox{\rm at}\ \rho<R_{\rm c}\\
   &\\ [-2mm]
    0,& \mbox{\rm at}\ \rho\ga R_{\rm c}.\\
    \end{array}\right.
\end{equation}
The amplitude of these dips in the background spectrum
should be very sensitive to its spatial fluctuations.
To reliably detect the effect of a decrease in
background brightness at the corresponding energies,
it is necessary to use extended (nearby) clusters.

In contrast, the increase in the background intensity below
$h\nu<60$ keV is associated with the scattered photons. These
are the photons of the radiation incident on the cluster from
all sides, least of all from its back side (the photons coming
from the back side are scattered at small angles and do not
contribute noticeably to the spectral
distortion). Figure\,\ref{fig:1scat} shows separately the
spectra of the distortions arising in the direct escape
radiation and scattered radiation in a cluster with a
temperature $kT_{\rm e}=5$ keV and a Thomson optical dept
%
%\noindent
$\tau_{\rm
  T}=1.2\times 10^{-2}$. It is clearly seen that all
``negative'' features (due to photoabsorption and the recoil
effect) are contained only in the direct escape spectrum; the
scattered photon spectrum is smooth and has no distinct
features. Formally, the scattered photons are also subject to
pho\-to\-absorption and are shifted to lower energies after
recoil (in secondary interactions with the gas), but these
effects are negligible due to the smallness of its optical
depth.
%------------------------------------------------------------------------------------------
\begin{figure*}[t]
%\centerline{\includegraphics[scale=0.66]{szxray_tau.ps}}
%\epsfxsize=0.51\textwidth
\begin{minipage}{0.49\textwidth}
  \includegraphics[width=1.02\textwidth]{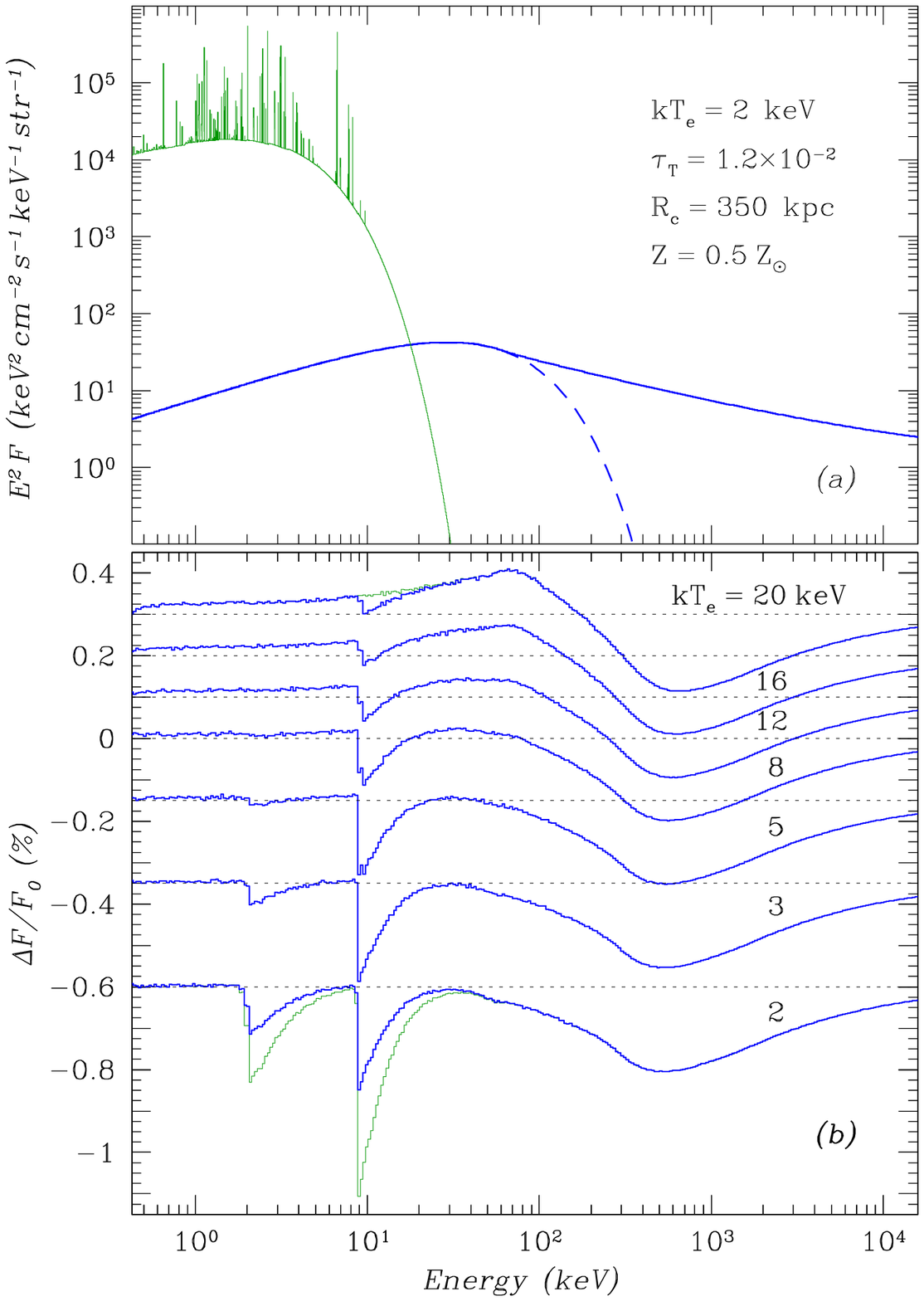}
%\epsffile{szxray_kt.ps}
\end{minipage}\begin{minipage}{0.49\textwidth}
  \includegraphics[width=1.02\textwidth]{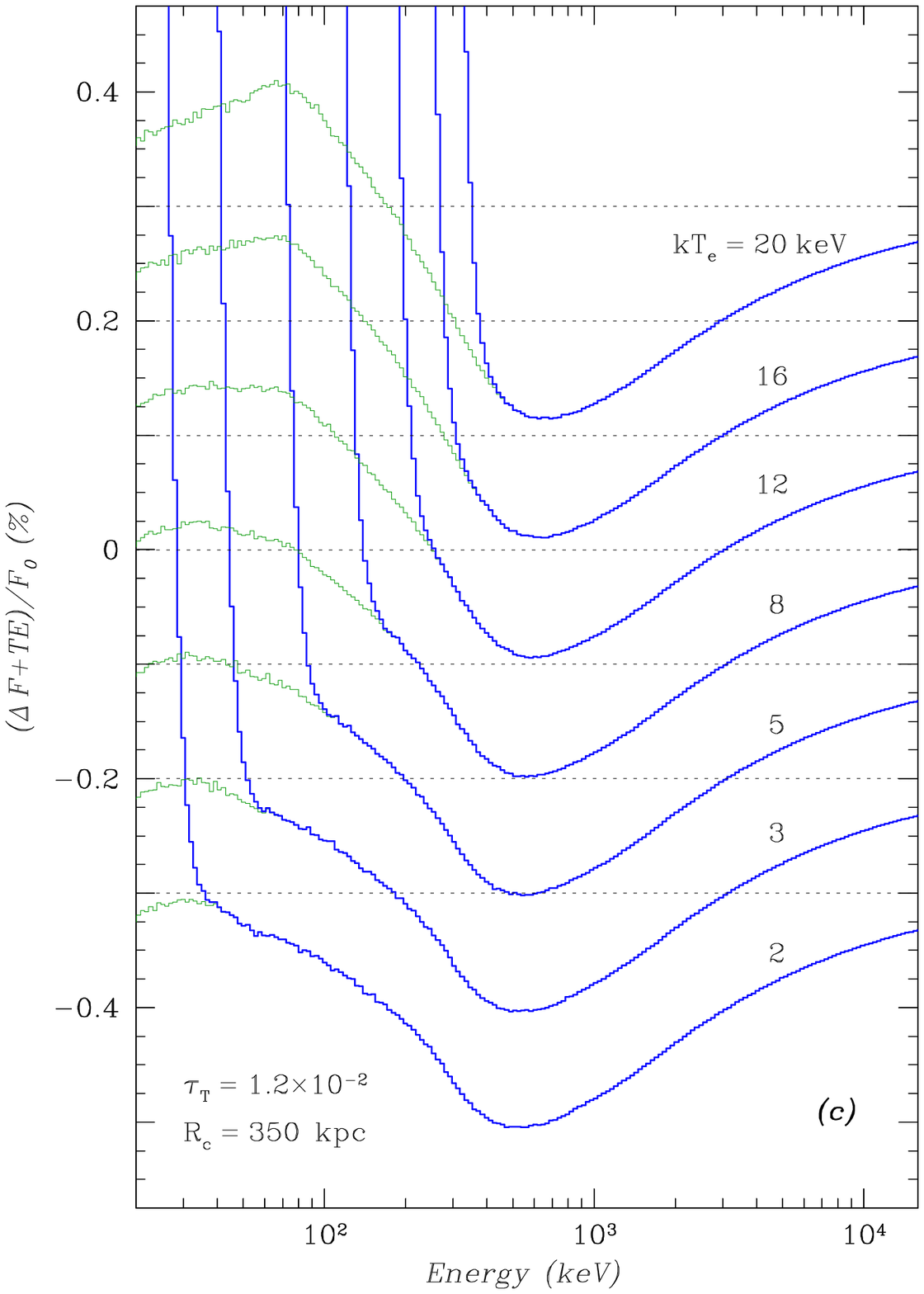}
%\epsffile{szxray_kt2.ps}
\end{minipage}   

\caption{\rm Same as Fig.\,\ref{fig:tau}, but for
  $\tau_{\rm T}=1.2\times10^{-2}$, $R_{\rm c}=350$ kpc ($N_{\rm
    e}=8.4\times10^{-3}\ \mbox{cm}^{-3}$), and various
  temperatures: (a) $kT_{\rm e}=2$ keV, (b, c) $kT_{\rm 
    e}=2,\ 3,\ 5,\ 8,\ 12,\ 16,\ \mbox{and}\ 20$ keV. The
  metallicity is $Z=0.5\ Z_{\odot}$. The green line on panel (b)
indicates our computation without photoabsorption  ($Z=0$) for 
  $kT_{\rm e}=20$ keV and with photoabsorption at metallicity
$Z=Z_{\odot}$ for $kT_{\rm e}=2$ keV. The background distortions
on panel (c) include the thermal plasma radiation (the thin
green lines --- without this radiation). \label{fig:kt}}
\vspace{-3mm} 
\end{figure*}
%---------------------------------------------------------------------------------------------

The contribution of various sky regions to the scattered
radiation spectrum is defined by the phase function, which in
the nonrelativistic limit $h\nu\ll m_{\rm e}c^2$ has a simple
form, $d\sigma_{\rm T}(\theta)=(3/8)\, \sigma_{\rm T}\,
(1+\cos\theta^2)\sin\theta\, d\,\theta$, where $\theta$ is the
scattering angle (and, in view of the symmetry of the phase
function, the photon arrival angle relative to the line of sight
toward the cluster). This is a smooth function. Clearly, the
photons responsible for the increase in background intensity in
the cluster due to the Doppler effect are collected from the
entire sky and, hence, these distortions should not be sensitive
to spatial background fluctuations.

%===========================================
\subsection{Dependence on the Gas Temperature}
\noindent
Figure\,\ref{fig:tau}a shows that the detection of distortions
in the background spectrum in the X-ray range $h\nu\la 60$ keV
will be greatly complicated due to the intrinsic thermal
radiation of the intergalactic gas.

The chances to detect the distortions increase for relaxed
clusters with a lower temperature. This is illustrated by
Fig.\,\ref{fig:kt}, in which the changes in the amplitude and
shape of the distortion spectrum are shown as a function of the
gas temperature. We again consider the model cluster with a
uniform density distribution within the radius $R_{\rm c}=350$
kpc and an optical depth along the line of sight passing through
the center $\tau_{\rm T}=1.2\times 10^{-2}$. In
Fig.\,\ref{fig:kt}a the gas temperature in the cluster was
assumed to be 2 keV. The figure gives a general idea of the
relationship between the background spectrum and the thermal
radiation spectrum of the cluster gas (just as
Fig.\,\ref{fig:tau}a, which gives such an idea for a cluster
with $kT_{\rm e}=5$ keV). Figure\,\ref{fig:kt}b shows the
relative distortions (in percent) arising in the background
spectrum after scattering and absorption in the gas of such a
cluster.  We considered various gas temperatures $kT_{\rm
  e}=2,\ 3,\ 5,\ 8,\ 12,\ 16,\ \mbox{and}\ 20$ keV.

The depth of the MeV dip in the background spectrum due to the
recoil effect after scattering by electrons is almost
independent of the gas temperature.  As would be expected, the
amplitude of other ``negative'' changes in the spectrum (the
dips due to photoabsorption) reach its maximum for a cold gas
with $kT_{\rm e}=2$ keV. The depth of the dip beyond the
threshold $h\nu\sim 9$ keV is $\sim0.25$\%. The amplitude of the
line at 2 keV also increases (to $\sim0.1$\%). These values
refer to the case where the metallicity in the gas is
$Z=0.5\,Z_{\odot}$.  The amplitude of the absorption lines
changes as the metallicity increases and decreases. \,In the
figure this change is indicated by the thin (green) lines for
$kT_{\rm e}=2$ keV ($Z=Z_{\odot}$) and $kT_{\rm e}=20$ keV
($Z=0$).

In contrast to the ``negative'' changes in the background
spectrum, the amplitude of the ``positive'' deviation (the
excess of background radiation due to the Doppler effect) is
maximal for the hottest gas with $kT_{\rm e}=20$ keV. At
energies $h\nu\sim 60-80$ keV a broad emission feature (line)
whose relative amplitude reaches $\sim0.1-0.15$\% is formed in
the corresponding spectrum of the background distortions partly
due to the Compton processes and partly due to the properties of
the background spectrum itself.
%---------------------------------------------------------------------------------------------
\begin{figure}[t]
\centerline{\includegraphics[width=1.04\linewidth]{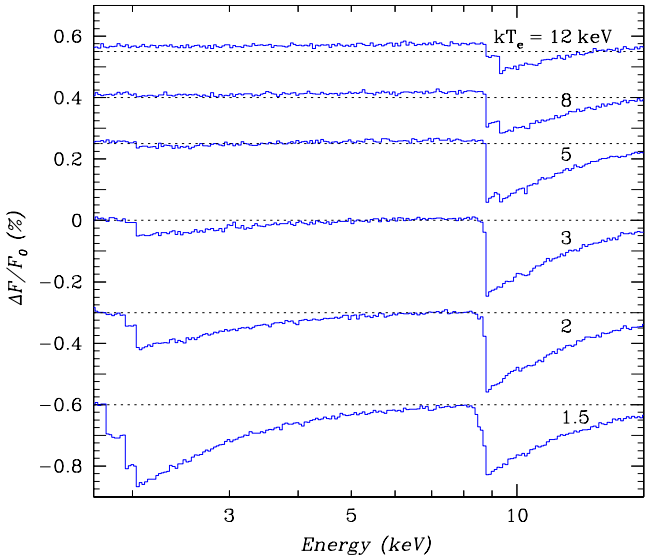}}
%\epsfxsize=1.04\linewidth
%\epsffile{szxray_lines.ps}
%\fbox{\rule{0cm}{6cm}\rule{0.97\linewidth}{0cm}}

\caption{\rm Formation of absorption lines in the X-ray
  background spectrum due to photoionization of iron and nickel
  ions in the hot gas of a galaxy cluster. A cluster with a
  uniform density, optical depth $\tau_{\rm
    T}=1.2\times10^{-2}$, gas temperatures $kT_{\rm
    e}=1.5,\ 2,\ 3,\ 5,\ 8,\ \mbox{and}\ 12$ keV, and
  metallicity $Z=0.5\,Z_{\odot}$. \label{fig:res.lines}}
\end{figure}
%---------------------------------------------------------------------------------------------

In Fig.\,\ref{fig:kt}c the thermal radiation of the
intergalactic gas was added to the background distortions (the
measurements are assumed to be performed toward the cluster
center). We see that the energy range that allows the Compton
distortions to be directly observed in the background spectrum,
without any illumination by the thermal gas radiation, turns out
to be sufficiently wide only for relaxed clusters with a gas
temperature $kT_{\rm e}\la5$ keV. For young clusters (or
clusters that have recently experienced a tidal effect from
another close cluster), where the gas has a higher temperature,
the lower boundary of the range admitting a direct observation
of the effect is shifted into the gamma-ray energy range
$h\nu\sim 400$ keV.

Note that the iron and nickel ions were assumed to be at rest
when computing the photoabsorption line profiles. This is
admissible, because the Doppler broadening and smearing related
to the thermal motion of ions, for example, for the profile of
the absorption jump at $h\nu_{\rm th}=9$ keV at typical gas
temperatures for clusters is only $\sim h\nu_{\rm th}\,(2
kT_{\rm e}/56\,m_{\rm p}c^2)^{1/2}\sim5\,(kT_{\rm
  e}/5\ \mbox{keV})^{1/2}$ eV. The resolution of the
computations presented in Figs.\,\ref{fig:tau} and \ref{fig:kt}
and most of the succeeding figures, $h\Delta \nu\sim
0.0345\,h\nu_{\rm th}\simeq 300$ eV, is much coarser. Even in
Fig.\,\ref{fig:res.lines}, which shows a detailed profile of
this line (with a resolution that is better by several times),
to demonstrate its complexity and multicomponent structure, the
thermal motion of iron and nickel ions could smooth only
slightly the sharpest features of the fine line structure.

Figure\,\ref{fig:res.lines} shows that the edge of the
absorption line near the threshold is strongly distorted even
without any thermal broadening. It has the shape of a more or
less regular ``step'' only in the case of $kT_{\rm e}=3$ keV. At
lower temperatures the beginning of both lines is shifted
leftward, the threshold turns into a semblance of a ``flight of
stairs'' consisting of several successive steps. At higher
temperatures additional steps appear on the right, shifting the
threshold of the lines by 300--400 eV to greater energies. On
reaching $kT_{\rm e}=8$ keV, the absorption line with the
threshold at $\sim2$ keV virtually disappears, the amplitude of
the hard line at $h\nu\sim9$ keV also decreases.

Table\,\ref{table:edge.keV} gives basic parameters of the line
with the threshold at $\sim 9$ keV, including its equivalent
width $W_{\rm X}$.  Since this line is formed in a region of
enhanced background flux (through the Compton continuum forming
in the cluster due to the Doppler effect after scattering by
electrons), its equivalent width $W_{\rm
  X}=\int^\infty_{\nu_{\rm th}}\,(F_{\rm c}-F_{\nu})/F_{\rm
  c}\,d\,h\nu$ was determined relative to this enhanced level
$F_{\rm c}(\nu)$ (the background photon spectrum distorted in
the cluster obtained in the limit $Z\rightarrow 0$). The table
presents the dependence of line parameters on the gas
temperature and metallicity.  As the temperature rises, the
effective threshold of the line $h\nu_{\rm th}$ and the energy
of the deepest point of its profile $h\nu_{\rm X}$ are shifted
by 400--500 eV to higher energies. The cause can be understood
from Fig.\,\ref{fig:res.lines} --- a new step related to the
absorption of photons by more strongly ionized iron and nickel
ions appears in the structure of the left line edge. For the
same reason, the line width $h\Delta\nu_{\rm X}$ (FWHM ), which
slightly increases with temperature probably due to a change in
the shape of the Compton continuum, decreases abruptly by
$\sim500$ eV on reaching $kT_{\rm e}=12$ keV. At this time the
step height reaches half the absorption line depth.
%---------------------------------------------------------------------------------------
\begin{table}[t]
  \caption{Parameters of the photoabsorption line with the
    threshold at $\sim 9$ keV in the X-ray background spectrum
    toward a cluster with a uniform gas
    density\aa\label{table:edge.keV}}

  \vspace{-2mm}

\begin{center} \small
  \begin{tabular}{c|c|c|c|c}\hline\hline
   \ $kT_{\rm e},$ \ \ \ & \ \ \ $h\nu_{\rm th}$\bb \ \ &
    \ \ \ $h\nu_{\rm X}$\cc \ \ &
   \  \ \ $h\Delta\nu_{\rm X}$\dd&$W_{\rm X}$\ee\\
    & & & & \\ [-3mm]
    \hline
      & & & & \\ [-2.5mm] 
   keV& keV&  keV& keV&eV\\ \hline
   \multicolumn{5}{c}{} \\ [-3mm]
    \multicolumn{5}{c}{$Z=0.5\,Z_{\odot}$}\\ \hline
    & & & & \\ [-2mm] 
    $1.5$   &~~8.29\ff&~~8.83\ff&2.98&12.7\\
    $2\,\,$  &8.38&8.83&3.06&14.0\\ 
    $3\,\,$  &8.58&8.83&3.18&13.6\\ 
    $5\,\,$  &8.58&8.83&3.69&11.9\\ 
    $8\,\,$  &8.68&9.35&3.91&\ 8.4\\ 
    $12\,$  &8.68&9.35&3.43&\ 5.3\\ 
    \hline
    \multicolumn{5}{c}{} \\ [-3mm]
    \multicolumn{5}{c}{$Z=1.0\,Z_{\odot}$}\\ \hline
    & & & & \\ [-2mm]
    $1.5$   &8.29&8.83&3.03&25.5\\ 
    $2\,\,$  &8.38&8.83&3.08&28.1\\ 
    $3\,\,$  &8.58&8.83&3.14&27.4\\ 
    $5\,\,$  &8.58&9.35&3.70&23.9\\ 
    $8\,\,$  &8.68&9.35&3.86&17.0\\ 
    $12\,$  &8.68&9.35&3.38&10.7\\  
\hline
\multicolumn{5}{l}{}\\ [-1mm]
\multicolumn{5}{l}{\aa\  $\tau_{\rm T}=1.2\times10^{-2}$.}\\
\multicolumn{5}{l}{\bb\ The threshold energy.}\\  
\multicolumn{5}{l}{\cc\ The energy of the deepest point of the line.}\\  
\multicolumn{5}{l}{\dd\ FWHM.}\\  
\multicolumn{5}{l}{\ee\ The equivalent width.}\\
\multicolumn{5}{l}{\ff\ The value reflects the resolution of our computation.}
\end{tabular}
\end{center}
\vspace{-5mm} 
\end{table}
%---------------------------------------------------------------------------------------

%===========================================
\subsection{The Model with a $\beta$ Density Distribution}
\noindent
Peripheral observations of the gas in clusters with
a real (decreasing with radius) density distribution
could also have a certain advantage in combatting
the thermal radiation. Indeed, the thermal radiation
intensity is proportional to $N_{\rm e}^2$, while the distortions
due to scattering by electrons are proportional to $N_{\rm e}$.
Accordingly, the contribution of the thermal radiation
should drop to the cluster edge faster than that of the
scattered one (Zel'dovich and Sunyaev 1982).

The action of this effect is demonstrated by
Fig.\,\ref{fig:tau}c, in which the distortions of the background
spectrum (including the thermal plasma radiation) are
represented by the thick (blue) lines for various impact
parameters $\rho$ from the cluster center. For comparison, the
thin (green) line indicates the spectrum of the distortions that
should be observed toward the center.  In the case of large
$\rho$ the lines intersect, suggesting that the contribution of
the thermal radiation vanishes at lower energies than those for
the observations toward the cluster center. The spectra shown in
this figure were computed for a $\beta$ gas density distribution
(Cavaliere and Fusco-Femiano 1976), which agrees satisfactorily
with the observed X-ray brightness distribution of many clusters
(Arnaud 2009),
\begin{equation}\label{eq:king}
N_{\rm e}=N_{\rm c}\left(1+\frac{R^2}{R_{\rm
    c}^2}\right)^{-3\beta/2}. 
\end{equation}
It follows from observations (Jones and Forman 1984) that for
most clusters $\beta\simeq 2/3$. At such $\beta$ the cluster
surface emission measure $EM(\rho)=2 \int_0^{\infty} N_{\rm
  e}^2(\rho)\,d\,l$, defining the thermal radiation intensity,
and the Thomson optical depth along the line of sight $\tau_{\rm
  T}(\rho)=2\sigma_{\rm T}\int_0^{\infty}N_{\rm e}(\rho)\,d\,l$,
defining the amplitude of the spectral distortions due to
scattering and absorption, at an impact parameter $\rho$ are,
respectively,
\begin{equation}\label{eq:kingme}
  EM(\rho)=\frac{\pi}{2}\left(1+\frac{\rho^2}{R_{\rm
      c}^2}\right)^{-3/2}\, N_{\rm c}^2\,R_{\rm c}
\end{equation}
and
\begin{equation}\label{eq:kingtau}
  \tau{\rm}_{\rm T}(\rho)=\pi\ \left(1+\frac{\rho^2}{R_{\rm
      c}^2}\right)^{-1/2}\, \sigma_{\rm T}\, N_{\rm c}\,R_{\rm c}.
\end{equation}
The gas emission measure $EM$ in a real cluster is seen to drop
with increasing $\rho$ much faster than the optical depth
$\tau_{\rm T}$. Therefore, at large $\rho$ the thermal radiation
should cease to hinder the background observations at lower
energies than in the observations toward the cluster
center. This is illustrated by Fig.\,\ref{fig:tau}c, based on
Eqs. (\ref{eq:kingme}) and (\ref{eq:kingtau}). Unfortunately, the
extension of the range favorable for the observation of
background distortions after scattering turns out to be
moderately large, while the amplitude of the effect proper, in
turn, decreases quite rapidly with increasing $\rho$.
%---------------------------------------------------------------------------------------------
\begin{figure}[t]
\centerline{\includegraphics[width=1.04\linewidth]{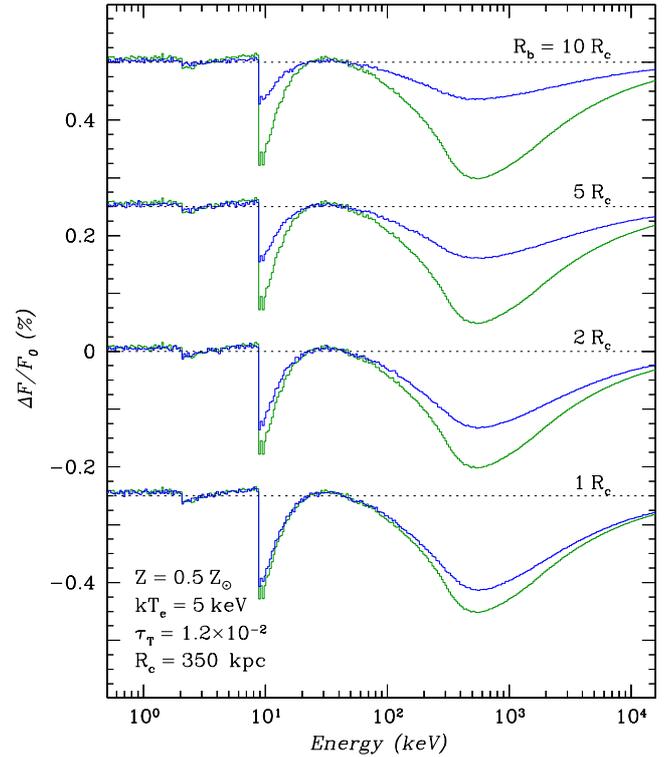}}
%\epsfxsize=1.04\linewidth
%\epsffile{szxray_z3.ps}
%\fbox{\rule{0cm}{6cm}\rule{0.97\linewidth}{0cm}}
\caption{\rm Comparison of the relative background distortions
  arising in a ``real'' galaxy cluster (thick blue lines) with a
  $\beta$ gas density distribution (Eq. (\protect\ref{eq:king}))
  and a cluster with a uniform density distribution (thin green
  line). The gas in the clusters has the same Thomson optical
  depths $\tau_{\rm T}=1.2\times10^{-2}$ (along the line of
  sight through the cluster center), core radii $R_{\rm c}=350$
  kpc, electron temperatures $kT_{\rm e}=5$ keV, and
  $Z=0.5\,Z_{\odot}$. We considered the cases with various break
  radii $R_{\rm b}$ of the $\beta$ profile. \label{fig:king}}
\end{figure}
%---------------------------------------------------------------------------------------------

In the computation whose result is used here, as before, the
intergalactic gas temperature was taken to be $kT_{\rm e}=5$
keV, the cluster core radius is $R_{\rm c}=350$ kpc, and the
Thomson optical depth of the gas along the line of sight passing
through the cluster center is $\tau_{\rm T}=1.2\times
10^{-2}$. The distortions in the background spectrum toward the
center (at $\rho=0$) were computed by the Monte Carlo method by
assuming the density profile to break at the ``outer'' radius
$R_{\rm b}=2\, R_{\rm c}$.

The dependence of the results of our computations on $R_{\rm b}$
is investigated in Fig.\,\ref{fig:king}. The spectra of the
background distortions arising in clusters with a $\beta$ gas
density distribution, the same temperatures $kT_{\rm e}$ and
optical depths along the line of sight toward the center
$\tau_{\rm T}$, but different break radii of the $\beta$ density
profile $R_{\rm b}=1,\ 2,\ 5,\ \mbox{and}\ 10\,R_{\rm c}$ are
shown here. For comparison, the thin lines indicate our
computations of the background distortions in a cluster with a
uniform density distribution (with the same $kT_{\rm e}$,
$R_{\rm c}$, and $\tau_{\rm T}$).

A change in $R_{\rm b}$ slightly changes the optical depth of
the gas $\tau_{\rm T}(\rho=0,R_{\rm b})$ in a real cluster
relative to $\tau_{\rm T}(\rho=0)=\pi \sigma_{\rm T} N_{\rm c}
R_{\rm c}$ following from Eq.\,(\ref{eq:kingtau}) (derived in
the limit $R_{\rm b}\rightarrow \infty$). Nevertheless, it is
appropriate to compare the clusters of equal optical depths for
the pure effect of different cluster geometries to be seen. For
this purpose, in the computations in Fig.\,\ref{fig:king} the
central density $N_{\rm c}$ of the cluster profile in
Eq.\,(\ref{eq:king}) was multiplied by $0.5\,\pi/\mbox{\rm
  arctan}(R_{\rm b}/R_{\rm c})$. With or without this correction,
the model clusters considered, of course, cannot be deemed
identical, if only because the cluster gas mass increases
noticeably with $R_{\rm b}$ from $M_{\rm
  g}=3.0\times10^{13}\ M_{\odot}$ for $R_{\rm b}=1\,R_{\rm c}$
to $M_{\rm g}=6.4\times10^{14}\ M_{\odot}$ for $R_{\rm
  b}=10\,R_{\rm c}.$

Figure\,\ref{fig:king} shows that the cluster with a real
density distribution with $R_{\rm b}=1\ R_{\rm c}$ leads to
virtually the same background distortions in amplitude and shape
of the energy dependence as does the cluster with a uniform
density distribution. It can also be seen from the figure that
even despite the increase in the mass of the cluster with a real
density distribution with increasing $R_{\rm b}$, the amplitude
of the background distortions decreases rapidly in this
case. This behavior can be explained by taking into account the
fact that the Thomson optical depth of the intergalactic gas
averaged over the visible area of the cluster with a $\beta$
profile, $<\!\tau_{\rm T}\!>$, decreases with increasing $R_{\rm
  b}.$ Indeed, by integrating the optical depth from
Eq.\,(\ref{eq:kingtau}) over the area $2\pi \int^{R_{\rm b}}_0
\tau_{\rm T}(\rho)\rho\,d\,\rho$ and normalizing to $\pi R_{\rm
  b}^2$, we find
\begin{equation}\label{eq:kingtauav}
  <\!\tau_{\rm T}(R_{\rm b})\!>=2\tau_{\rm T}\frac{R_{\rm c}^2}{R_{\rm b}^2}\left[\left(1+\frac{R_{\rm b}^2}{R_{\rm c}^2}\right)^{1/2}\!-1\right]
\end{equation}
$\simeq0.83\ \tau_{\rm T}$ at $R_{\rm b}=1\, R_{\rm c}$ and
$\simeq0.18\ \tau_{\rm T}$ at $R_{\rm b}=10\ R_{\rm c}.$

Note that we compute the average spectrum of the
background distortions in the cluster by the Monte
Carlo method. To obtain the background distortions
toward its center ($\rho=0$)  presented in Fig.\,\ref{fig:tau}c, their
amplitude was properly corrected for the above decrease
in optical depth when averaged over the visible
cluster area $\pi (2\,R_{\rm c})^2$.
%---------------------------------------------------------------------------------------------
\begin{figure}[t]
  \vspace{-3mm}
\centerline{\includegraphics[width=1.04\linewidth]{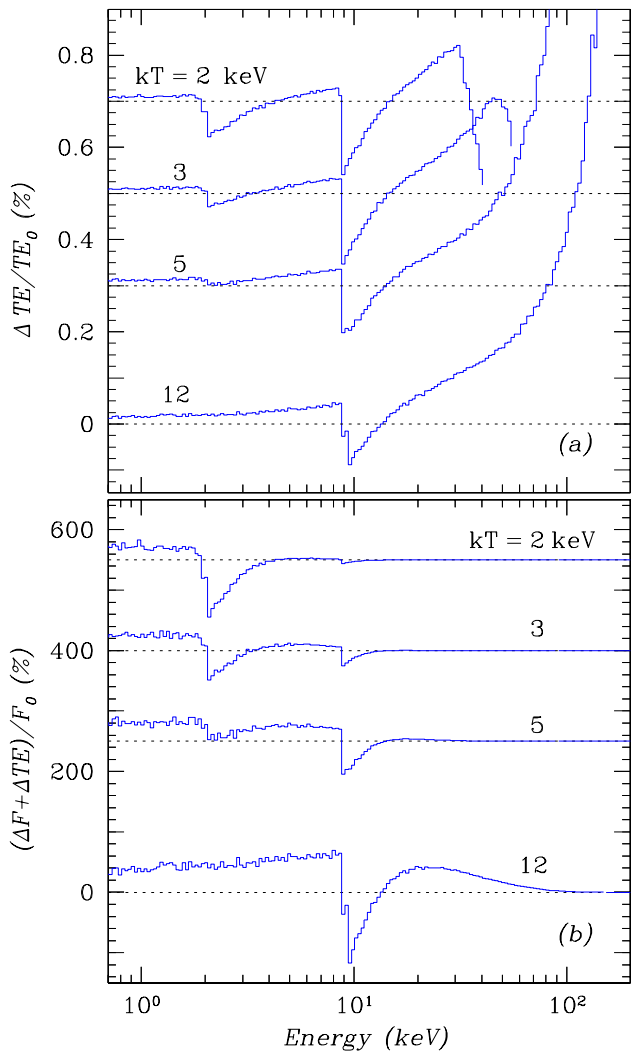}}
%\epsfxsize=1.04\linewidth
%\epsffile{szxray_tbr.ps}
%\fbox{\rule{0cm}{8cm}\rule{0.97\linewidth}{0cm}}
\caption{\rm Formation of absorption lines in the thermal
  radiation spectrum of the cluster gas due to the ionization of
  iron ions: relative to (a) the thermal radiation itself and
  (b) the cosmic background radiation. A cluster with a uniform
  density, radius $R_{\rm c}= 350$ kpc, $\tau_{\rm
    T}=1.2\times10^{-2}$, $Z=0.5\, Z_{\odot},$ and various gas
  temperatures $kT_{\rm e}$. \label{fig:gas.edge}}
\vspace{-3mm}
\end{figure}
%---------------------------------------------------------------------------------------------

>From the viewpoint of effective detection of the background
distortions due to the interaction with the cluster gas,
Fig.\,\ref{fig:king} clearly indicates that the cluster
observations by a telescope with an aperture (angular
resolution) covering the central part of the cluster with a
radius $\la 2 R_{\rm c}$. A similar conclusion can be drawn from
an analysis of the background distortions arising in a cluster
with the density distribution predicted by the
Navarro-Frenk-White model (hereafter NFW, Navarro et
al. 1997). Such a cluster is analyzed in the Appendix.

%===========================================
\subsection{Spectral Distortions of the Intrinsic Gas Radiation}
\noindent
The detection of the background distortions associated
with its interaction with the hot gas of galaxy
clusters is complicated not just by the presence of
intense intrinsic gas radiation. In turn, distortions
whose amplitude exceeds noticeably the relative amplitude
of the background distortions appear in the
spectrum of this thermal radiation.

Indeed, the Raymond-Smith code, along with other codes used to
compute the bremsstrahlung and recombination radiation spectrum
of an optically thin plasma, suggests that the optical depth of
the plasma $\tau_{\rm T}\rightarrow 0$ and, therefore, takes
into account only the collisional processes and disregards the
ionization of the iron-group elements by the intrinsic plasma
radiation. It also disregards the Compton scattering of the
plasma radiation inside the cluster. Meanwhile, as the results
of our computations presented in this paper show, the optical
depth of the cluster gas is enough for strong absorption lines
at $h\nu\ga 2$ and $9$ keV and other features to be formed in
the spectrum when the background radiation passes through
it. Obviously, such spectral distortions should also appear in
the intrinsic gas radiation.

Figure\,\ref{fig:gas.edge}a shows the results of our Monte Carlo
computations of such distortions (in \% to the thermal radiation
of the intergalactic gas) for a cluster with a uniform density
with $\tau_{\rm T}=1.2\times10^{-2}$, $kT_{\rm
  e}=2,\ 3,\ 5,\ \mbox{or}\ 12$ keV and $Z=0.5\, Z_{\odot}$. We
used the same code as that for the background distortion
computations, but the source of photons was assumed to be
uniformly distributed throughout the cluster and its radiation
spectrum was taken from our computations of the optically thin
plasma spectrum by the Raymond-Smith code for a cluster
temperature $kT_{\rm e}$ (only the continuum bremsstrahlung was
taken into account).
%---------------------------------------------------------------------------------------
\begin{table}[ht]
  \caption{Parameters of the total (including the thermal radiation
distortions) photoabsorption line with the threshold at $\sim 9$
   keV in the X-ray background spectrum toward a
cluster with a uniform gas density\aa\label{table:edge2.keV}}

  \vspace{-2mm}

\begin{center} \small
  \begin{tabular}{c|c|c|c|c}\hline\hline
    $kT_{\rm e},$& \ \ \ $h\nu_{\rm th}$\bb \ &
    \ \ \ $h\nu_{\rm X}$\cc \ &
    \ \ $h\Delta\nu_{\rm X}$\dd&$W_{\rm X}$\ee\\ \hline
   keV& keV&  keV& keV&keV\\ \hline
   \multicolumn{5}{c}{} \\ [-3mm]
    \multicolumn{5}{c}{$Z=0.5\,Z_{\odot}$}\\ \hline
    & & & & \\ [-2mm] 
    $2\,\,$  &8.19&8.93&1.13&0.11\\ 
    $3\,\,$  &8.48&8.93&1.33&0.61\\ 
    $5\,\,$  &8.48&8.93&2.16&2.22\\ 
    $8\,\,$  &8.48&9.57&2.89&3.59\\ 
    $12\,$  &8.48&9.57&2.77&3.72\\ %7.13\\  
    \hline
    \multicolumn{5}{c}{} \\ [-3mm]
    \multicolumn{5}{c}{$Z=1.0\,Z_{\odot}$}\\ \hline
    & & & & \\ [-2mm]
    $2\,\,$  &8.19&8.93&1.04&0.24\\ 
    $3\,\,$  &8.48&8.93&1.36&1.24\\ 
    $5\,\,$  &8.48&8.93&2.13&4.38\\
    $8\,\,$  &8.48&9.57&2.93&7.34\\
    $12\,$  &8.48&9.57&2.63&7.14\\  
\hline
\multicolumn{5}{l}{}\\ [-1mm]
\multicolumn{5}{l}{\aa\ $R_{\rm c}=350$ kpc, $\tau_{\rm T}=1.2\times10^{-2}.$}\\
\multicolumn{5}{l}{\bb\ The threshold energy.}\\  
\multicolumn{5}{l}{\cc\ The energy of the deepest point of the line.}\\  
\multicolumn{5}{l}{\dd\ FWHM.}\\  
\multicolumn{5}{l}{\ee\ The equivalent width.}
\end{tabular}
\end{center}
\vspace{-5mm} 
\end{table}
%---------------------------------------------------------------------------------------

We see that the distortions arising in the spectrum of the
thermal gas radiation resemble those in the background
spectrum. However, there is also a difference. The bremsstrahlung
spectrum abruptly breaks at energies $h\nu\ga kT_{\rm e}$. At the
same time, Comptonization shifts the photons upward along the
frequency axis, tending to form a Wien spectrum with a
characteristic break energy $\sim3kT_{\rm e}$. This process
explains the sharp increase in radiation intensity at energies
$h\nu\ga kT_{\rm e}$. The process is well known in X-ray
astronomy and is successfully used to explain the observed
spectra of accreting black holes (see, e.g., Shapiro et
al. 1976; Sunyaev and Titarchuk 1980).  The upward shift of the
photons along the frequency axis is determined by the Doppler
effect; after each scattering the average change in photon
frequency is $\Delta\,\nu/\nu\sim (kT_{\rm e}/m_{\rm
  e}c^2)^{1/2}$. A competing process is the recoil effect, which
in the nonrelativistic limit lowers the photon frequency after
scattering, on average, by $\Delta\,\nu/\nu\sim -h\nu/m_{\rm
  e}c^2$ (these estimates can be easily obtained from the
Kompaneets equation, see below). The processes are balanced at
\begin{equation}\label{eq:compton.peak}
  h\nu_*\sim (kT_{\rm e}\,m_{\rm e}c^2)^{1/2}\sim 32\,(kT_{\rm
    e}/2\ \mbox{\rm keV})^{1/2}\ \mbox{\rm keV}.
\end{equation}
Accordingly, for clusters with a low gas temperature $kT_{\rm
  e}\la2-3$ keV a broad emission feature centered at energies
$\sim 30-50$ keV is formed in the distortion spectrum
(Fig.\,\ref{fig:gas.edge}); for hotter clusters the distortions
grow with energy until the complete cutoff of the thermal
spectrum. Formula\,(\ref{eq:compton.peak}) to some extent also
explains the energy of the broad emission feature appearing in
the distortion spectrum of the background in hot galaxy clusters
(Fig.\,\ref{fig:kt}b).
 
The amplitude of the distortions in the thermal plasma radiation
spectrum is $\sim \tau_{\rm T}$ and accounts for fractions of
percent of the intensity of the spectrum itself (just as the
amplitude of the cosmic background distortions). However, since
the thermal radiation intensity in the X-ray range exceeds the
background intensity by two or three orders of magnitude, these
distortions are comparable to the background intensity.  When
the observed cluster radiation spectrum is fitted by the thermal
radiation model of an optically thin plasma, these distortions,
including the iron ion absorption lines, will not be subtracted
and will lead to a noticeable enhancement of the distortions
formed directly in the background spectrum. This is clearly seen
from Fig.\,\ref{fig:gas.edge}b, in which the background
distortions and the thermal gas radiation distortions are added
together and are given in percent relative to the background
spectrum. The background level is shown by dotted lines. The
amplitude of the distortions in the iron and nickel
photoabsorption lines reaches 100\% or more. Note that the
absorption line with the 9-keV threshold greatly weakens in the
distortion spectra of cold clusters with $kT_{\rm e}=2-3$ keV in
Fig.\,\ref{fig:gas.edge}b, although in Fig.\,\ref{fig:gas.edge}a
it has the largest amplitude among all of the clusters
considered relative to the thermal radiation spectrum. This is
due to the general rapid (exponential) drop in thermal radiation
at these energies from cold clusters. The line with the
threshold at $\sim2$ keV in them still reaches a
maximum. Table\,\ref{table:edge2.keV} gives the equivalent
widths of the effective (total) absorption line at 9 keV in the
background spectrum in clusters with various gas temperatures
and various iron abundances. Clearly, the intrinsic gas
radiation distortions are very large and at energies
$h\nu\la100$ keV exceed noticeably the background distortions. They
are very difficult to take into account.

Below we will assume that the question about the subtraction of
the thermal gas radiation spectrum, given its distortions due to
the finite optical depth, from the observed cluster spectrum has
somehow been solved. We will consider only the distortions
arising directly in the cosmic background. The distortions of
the intrinsic thermal gas radiation will be considered in detail
in a separate paper (Grebenev and Sunyaev 2020).

%xxxxxxxxxxxxxxxxxxxxxxxxxxxxxxxxxxxxxxxxxxxxxxxxxxxxxxxxxxxxxxx
\section*{ANALYTICAL ESTIMATES}
\noindent
When investigating the Compton scattering of the cosmic
microwave background radiation in the hot gas of a galaxy
cluster, important analytical estimates (Sunyaev and Zel'dovich
1980; Zel'dovich and Sunyaev 1982) were obtained by solving the
Kompaneets (1957) equation, which describes the photon energy
redistribution energy in the diffusion approximation.  The
validity of applying this equation to an optically thin gas
typical for clusters was tested and confirmed by Sunyaev
(1980). Similar quite interesting estimates can also be obtained
for the problem under consideration.
%---------------------------------------------------------------------------------------------
\begin{figure}[t]
\centerline{\includegraphics[width=1.03\linewidth]{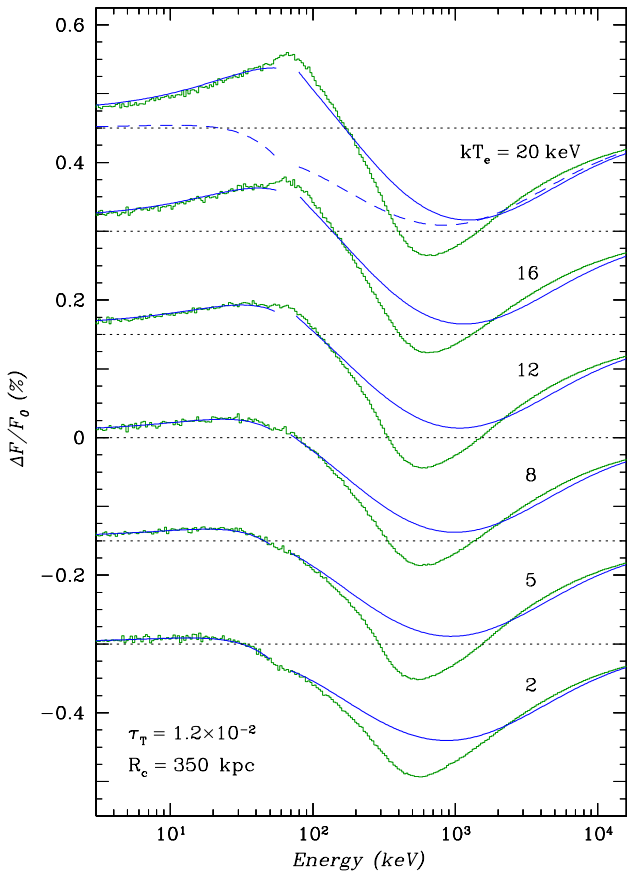}}
%\epsfxsize=1.03\linewidth
%\epsffile{szxray_komp.ps}
%\fbox{\rule{0cm}{10cm}\rule{0.97\linewidth}{0cm}}

\caption{\rm Comparison of the relative hard X-ray background
  distortions due to scattering in the hot cluster gas
  calculated from the analytical formulas\,(\ref{eq:relint}) and
  (\ref{eq:relintpl}) (solid thick blue lines) and by the Monte
  Carlo method (thin green lines). The photoabsorption by iron
  ions was disregarded. We considered a cluster with a uniform
  density distribution, core radius $R_{\rm c}=350$ kpc, optical
  depth $\tau_{\rm T}=1.2\times10^{-2},$ and various electron
  temperatures $kT_{\rm e}=2,\ 5,\ 8,\ 12,\ 16,\ \mbox{and}\ 20$
  keV. For $kT_{\rm e}=20$ keV the dashed line indicates the
  calculation including only the recoil effect after
  scattering.\label{fig:komp}}
\end{figure}
%---------------------------------------------------------------------------------------------

The Kompaneets equation with relativistic corrections (Cooper
1971; see also Arons 1971; Illarionov and Sunyaev 1972; Grebenev
and Sunyaev 1987), but at a moderately high electron temperature
($kT_{\rm e}\ll m_{\rm e}c^2$) can be represented as
\begin{equation}\label{eq:comp}\nonumber
  \frac{\partial F_{\nu}}{\partial\tau_{\rm c}}=\frac{h}{m_{\rm e}c^2}
  \frac{\partial}{\partial\nu}\left[
    \frac{\nu^4 \xi(T_{\rm e})}{1+\beta\nu+\gamma\nu^2}\left(\frac{F_{\nu}}{\nu^2}+
    \frac{kT_{\rm e}}{h}\frac{\partial}{\partial\nu}
    \frac{F_{\nu}}{\nu^2}\right)\right], 
  \end{equation}   
where $F_{\nu}$ is the intensity of the photon spectrum, $\tau_{\rm
  c}$ is the Thomson radial optical depth of the cloud,
$\beta=9\times10^{-3}\ \mbox{keV}^{-1},$
$\gamma=4.2\times10^{-6}\ \mbox{keV}^{-2},$ $$\xi(T_{\rm
  e})=1+\frac{5}{2}\frac{kT_{\rm e}}{m_{\rm e}c^2}.$$ We neglected the term responsible for the induced
scattering and retained only the term of the first order
in $kT_{\rm e}/m_{\rm e}c^2$ in $\xi(T_{\rm e})$ (see Cooper 1971). Substituting
the background intensity
$F_{\nu}(\nu)=S_{\nu}(\nu)/h\nu$ in the
form $F_{\nu}=A\nu^{-\alpha}
\exp{(-\nu/\nu_0)}$ (the part of the spectrum from Eq.\,(1)
corresponding to low energies $h\nu\la 60$ keV) into the right-hand side of this equation, for
the relative changes in the background we find 
\begin{equation}\label{eq:relint}
\frac{\Delta\,F_{\nu}}{F_{\nu}}= \frac{h\nu}{m_{\rm
    e}c^2}\frac{\tau_{\rm c}\ \xi(T_{\rm e})}{1+\beta\nu+\gamma\nu^2}
\left[1-\frac{kT_{\rm e}}{h\nu_0}+\right.
\end{equation}
\begin{equation}\nonumber
  \left. +\left(1-\alpha-\frac{\nu}{\nu_0}-\frac{\nu}{\nu_1}\right)
  \left(1-\frac{kT_{\rm e}}{h\nu_0}-
  \frac{kT_{\rm e}}{h\nu}\,(2+\alpha) \right) \right].
\end{equation}
Here, we introduce the notation $\nu_1$ for the function 
\begin{equation}\nonumber
  \nu_1(\nu)=\frac{1+\beta\nu+\gamma\nu^2}{\beta+2\gamma\nu}.
\end{equation}
The distortion spectrum for the harder, $h\nu\ga60$ keV,
(power-law) part of the spectrum can be found from
Eq.\,(\ref{eq:relint}) by passing to the limit
$\nu_0\rightarrow\infty$:
\begin{equation}\label{eq:relintpl}
\frac{\Delta\,F_{\nu}}{F_{\nu}}= \frac{h\nu}{m_{\rm
    e}c^2}\frac{\tau_{\rm c}\ \xi(T_{\rm e})}{1+\beta\nu+\gamma\nu^2}\times
\end{equation}
\begin{equation}\nonumber
\times  \left[1+\left(1-\alpha-\frac{\nu}{\nu_1}\right)
  \left(1-\frac{kT_{\rm e}}{h\nu}\,(2+\alpha) \right) \right].
\end{equation}
In the limit $\beta\nu\ll 1$ and $\nu\ll\nu_0$ Eq.\,(\ref{eq:relintpl}) gives
\begin{equation}\nonumber
\frac{\Delta\,F_{\nu}}{F_{\nu}}= (2+\alpha)\tau_{\rm c} \frac{kT_{\rm e}\xi(T_{\rm
  e})}{m_{\rm e}c^2} \left[\alpha-1+\frac{2-\alpha}{2+\alpha}\frac{h\nu}{kT_{\rm e}} \right].
\end{equation}
Accordingly, to a first approximation, for $h\nu\la kT_{\rm e}$,
the amplitude of the effect is proportional to the Compton
parameter $y_{\rm C}=\tau_{\rm c} \xi (T_{\rm e}) kT_{\rm e}/
m_{\rm e}c^2$.  In the opposite limit $h\nu\gg kT_{\rm e}$ the
formula for the distortions takes the form
\begin{equation}\nonumber
\frac{\Delta\,F_{\nu}}{F_{\nu}}= - \tau_{\rm c}\ \xi(T_{\rm e}) \frac{h\nu}{m_{\rm
    e}c^2}\frac{\alpha-2+\nu/\nu_1}{1+\beta\nu+\gamma\nu^2},
\end{equation}
i.e., the amplitude of the effect is proportional to $\tau_{\rm c}.$

The thick solid (blue) lines in Fig.\,\ref{fig:komp} indicate
the results of applying Eqs.\,(\ref{eq:relint}) and
(\ref{eq:relintpl}) to a cluster with a radial optical depth
$\tau_{\rm c}=6\times10^{-3}$ (which corresponds to the optical
depth along the line of sight passing through the center
$\tau_{\rm T}=1.2\times10^{-2}$) and electron temperatures
$kT_{\rm e}=2,\ 5,\ 8,\ 12,\ \mbox{and}\ 20$ keV. At $h\nu>60$ keV we used a
superposition of the solutions (\ref{eq:relintpl}) for the
spectra with different photon indices in accordance with the
model of the background spectrum (Eq.\,(1)). Since the background
spectrum was fitted by different functions at low and high
energies, its derivative can have a discontinuity at 60 keV. In
the distortion spectra calculated from the approximate formulas
(\ref{eq:relint}) and (\ref{eq:relintpl}), a jump or break is
observed near this energy for many temperatures. Therefore, the
analytical solution in the narrow region $\pm10$ keV near 60 keV
is not shown in the figure for clarity.  For clarity, we also
disregard the photoabsorption of background photons by strongly
ionized iron ions.  This process can be easily included in the
analytical solution (see Grebenev and Sunyaev 1987).
%---------------------------------------------------------------------------------------------
\begin{figure*}[t]
%\centerline{\includegraphics[scale=0.66]{szxray_tau.ps}}
%\epsfxsize=0.51\textwidth
\begin{minipage}{0.49\textwidth}
  \includegraphics[width=1.02\textwidth]{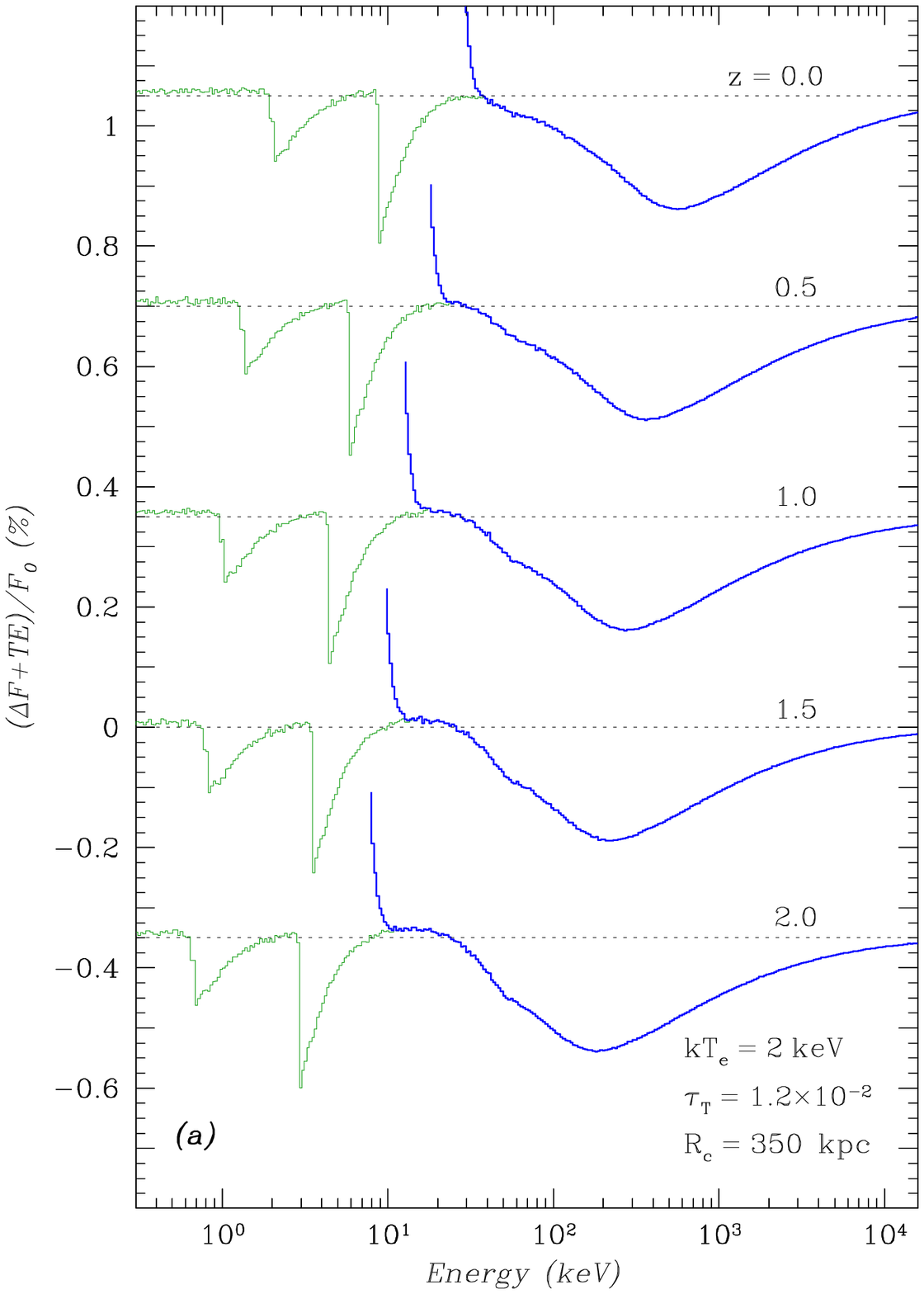}
  %\epsffile{szxray_z1.ps}
\end{minipage}\begin{minipage}{0.49\textwidth}
  \includegraphics[width=1.02\textwidth]{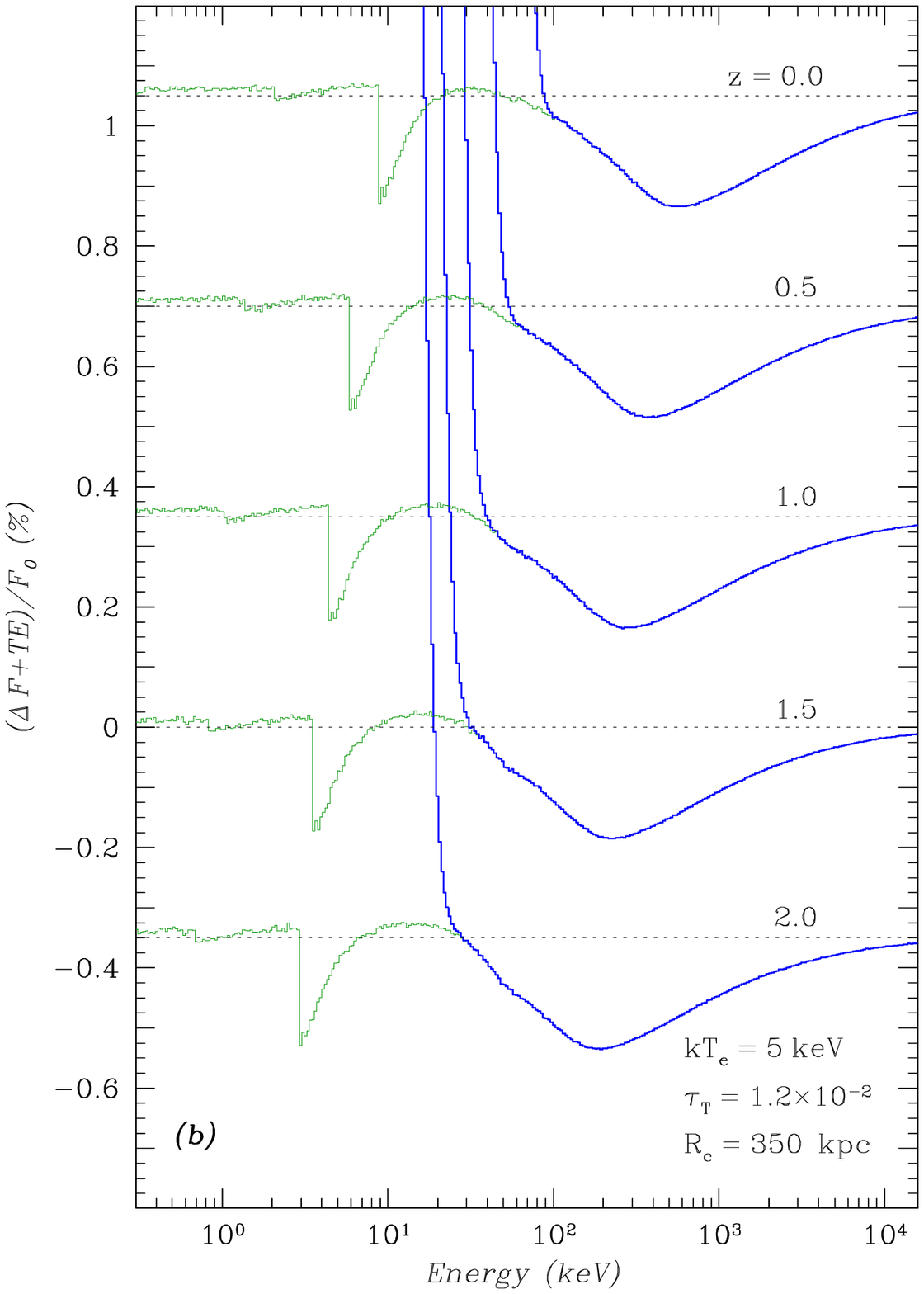}
  %\epsffile{szxray_z2.ps}
\end{minipage}

\caption{\rm Expected cosmic background (its intensity)
  distortions after scattering and absorption in the hot cluster
  gas versus its redshift $z$ (thin green lines). Clusters with
  a uniform density, optical depth $\tau_{\rm
    T}=1.2\times10^{-2},$ and core radius $R_{\rm c}=350$
  kpc. The gas temperature is $kT_{\rm e}=$ (a) $2$ and (b) $5$
  keV, $Z=0.5\,Z_{\odot}$. The thick (blue) lines indicate the
  background distortions including the thermal plasma radiation
  (shown incompletely on panel (a)).\label{fig:z2}}
\end{figure*}
%---------------------------------------------------------------------------------------------

The thin (green) lines in Fig.\,\ref{fig:komp} indicate the
results of our Monte Carlo computations of the distortions in
the background spectrum. The computations were performed for the
same cluster parameters as those for the analytical solution and
allow them to be compared.  However, they differ not only by the
method of solution, but also by the boundary conditions: for the
Kompaneets equation the isotropic background source is located
at the gas cloud center, while for the Monte Carlo method the
background radiation penetrates the cloud from outside. Note
that an indistinct feature, which, obviously, is associated with
the piecewise continuous fit of the background spectrum
(Eq.\,(1)), is also observed near 60 keV in many of the spectra
computed by the Monte Carlo method.

On the whole, it can be said that the calculation based on
Eqs.\,(\ref{eq:relint}) and (\ref{eq:relintpl}) correctly
reproduces the spectral shape of the numerically computed
distortions, although it smoothes the deepest part of the
absorption feature at energies $h\nu\sim 500-600$ keV arising
due to the recoil effect.  It is possible that a better
coincidence could be achieved here by numerically integrating
the relativistic kernel of the kinetic equation for Compton
scattering (Sazonov and Sunyaev 2000), but in this case an
expression for the spectrum could not be obtained
explicitly. The presented formulas give quite reasonable
estimates of the distortions in the background spectrum. The
analytical solution also allows the nature of the various
components in the distortion spectrum to be easily
clarified. For example, the dashed line in Fig.\,\ref{fig:komp}
indicates our calculation of the distortions in the limit
$T_{\rm e}\rightarrow 0$, i.e., arising in the background
spectrum due to the recoil effect. Clearly, almost the entire
excess of radiation at $h\nu\la150$ keV and the shape of the
left edge of the absorption feature at $h\nu\ga150$ keV are
associated with the Maxwellian velocities of electrons and are
formed due to the Doppler effect.

%===========================================
\section*{DEPENDENCE ON $Z$}
\noindent
One of the remarkable properties of the effect of microwave
background radiation scattering by the hot gas of galaxy
clusters is its independence of the redshift $z$. Indeed, no
matter how far the cluster is, the distortions are observed in
its present-day well measured spectrum characterized by a
temperature $T_{\rm r}= 2.7$ K. Although the temperature $T_{\rm
  r}$ was a factor of $(1+z)$ higher at the time of its
interaction with the cluster, this does not manifest itself in
any way, because the equation describing the Doppler spectral
distortions depends only on $x=h\nu/kT_{\rm r}$ (Zel'dovich and
Sunyaev 1982), i.e., it is invariant in $z$.

The situation with the X-ray background scattering is
different. Here all distortions are also observed in the
present-day spectrum, although they are formed in the background
spectrum at the cluster redshift $z$. However, apart from the
Doppler ones, among them there are the distortions that arise
due to photoabsorption and the recoil effect at quite specific
energies --- the absorption thresholds $h\nu\sim2$ and $9$ keV
and at $h\nu\sim500$ keV. In the spectrum observed at $z>0$
these features turn out to be shifted to low
energies. Figure\,\ref{fig:z2} presents the spectra of the
background distortions that should be recorded from clusters
with the same parameters ($\tau_{\rm T}=1.2\times10^{-2}$,
$R_{\rm c}=350$ kpc, and $kT_{\rm e}=2$ keV in
Fig.\,\ref{fig:z2}a or $5$ keV in Fig.\,\ref{fig:z2}b), but
located at different redshifts. The spectra were obtained by
recalculating the present-day background spectrum (Eq.\,(1)) to
the corresponding $z$ using standard formulas (see, e.g.,
Zel'dovich and Novikov 1975), computing the distortions there,
and then recalculating the spectrum back to $z=0$\footnote{In
  our calculations we adopted the standard $\Lambda$CDM
  cosmological model with $\Omega_{\rm M}=0.3,$
  $\Omega_{\Lambda}=0.7,$ and $H=70\ \mbox{\rm km
    s}^{-1}\ \mbox{Mpc}^{-1}$.}.  Both (photoabsorption and
recoil effect) lines are seen to be actually greatly shifted
leftward; at $z=2$ the absorption line is at $\sim3$ keV, while
the minimum of the feature related to the recoil effect is at
$\sim200$ keV. However, the case of $z=2$ is, in a sense,
extreme --- the clusters are subject to strong evolution and at
$z\sim2$ they (especially massive clusters) were much fewer than
those observed now. Note that the Doppler spectral distortions
(the excess of radiation), which clearly manifest themselves at
low energies in Fig.\,\ref{fig:z2}b, are almost independent of
$z$, like the microwave background distortions.
%---------------------------------------------------------------------------------------------
\begin{figure}[t]
\hspace{-2mm}\includegraphics[width=1.04\linewidth]{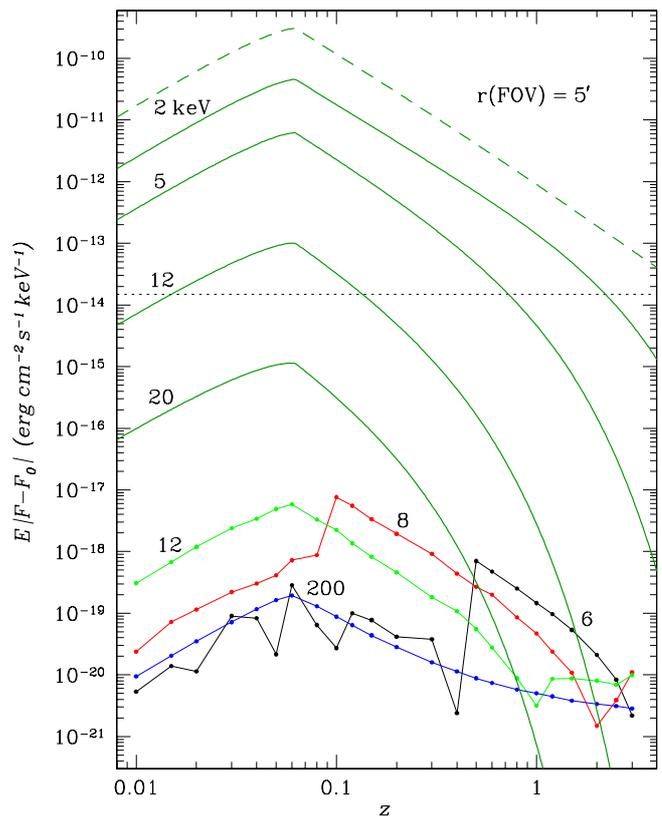}
%\epsffile{szxray_z.ps}
%\fbox{\rule{0cm}{9cm}\rule{0.97\linewidth}{0cm}}

\caption{\rm Variations in the spectral flux (recorded within
  the telescope aperture with a $5$\arcmin\ radius) of thermal
  X-ray radiation from the hot gas in a galaxy cluster (dark
  green lines) and the absolute amplitude of the background
  distortions due to Compton scattering and photoabsorption in
  this gas (lines with dots) with $z$. The energy $h\nu$ for
  which the flux is given is specified near each curve. The
  dashed line indicates the change in the integrated flux of
  thermal radiation. The dotted line indicates the X-ray
  background flux at 5 keV. The cluster gas has an optical depth
  for Thomson scattering $\tau_{\rm T}=1.2\times10^{-2}$, core
  radius $R_{\rm c}=350$ kpc, and temperature $kT_{\rm e}=2$
  keV.\label{fig:z}}
\end{figure}
%---------------------------------------------------------------------------------------------

%---------------------------------------------------------------------------------------------
\begin{table*}[t] 
\caption{Parameters of the individual clusters selected to
  estimate the effect of scattering and absorption of the X-ray
  background in their hot intergalactic gas$^*$\label{table:clusters}}

\vspace{3mm}
%\begin{center}
\small
\hspace{-1mm}\begin{tabular}{l|@{\,}l@{\,}|l|c|@{\,}c|c|c|@{\,}c|c|c|c|c|c|l@{}}\hline
\multicolumn{2}{c|}{}             &                     &      & &      &       &           & & &       &       &&\\ [-3mm]
\multicolumn{2}{c|}{Cluster name\aa}&\multicolumn{1}{c|}{$z$}&$R_{500},$\bb&$R_{\rm c},$&
$\theta_{\rm c},$&$kT_{\rm e},$& $N_{\rm c}$\cc&$M_{\rm g}$&$M_{500}$\bb&$Y_{\rm SZ}$\dd&Z\ee,&$\tau_{\rm T}$,\ff&
\!Reference\g2\\ \cline{1-2} \cline{9-10}
\multicolumn{2}{c|}{}             &                     &      & &      &       &           & \multicolumn{2}{c|}{} &       &       &&\\ [-3mm]
\multicolumn{1}{c|}{main}&\multicolumn{1}{c|}{alternative}&&kpc&kpc&\arcmin&keV&&
\multicolumn{2}{c|}{$10^{14}\ M_{\odot}$}& &$Z_{\odot}$&$10^{-3}$&\\ \hline
 & & & & & & & & & & & &&\\ [-3.5mm]
 \multicolumn{1}{c}{1}&\multicolumn{1}{c}{2}& \multicolumn{1}{c}{3}&
 4    & 5&  6    &  7     &8  &  9    & 10      &  11     &  12
 &13&\  \ \  14\\  \hline 
 & & & & & &  & & & & & & &\\ [-3.6mm]
AT\,J0102-4915&El Gordo&0.870             &\underline{950}&\underline{270}&0.75&14.5&\underline{8.9}&2.2 &11.9&32.0& 0.2&15.2  &1,2,3\\
A\,426&Perseus&0.018&\underline{1400}&280&12.8&6.0&4.6&2.0& \underline{8.0}& \underline{11.9}&0.5&\underline{8.3}&4,5,6\\
ST\,J0615-5746&P\,G266.6-27.3\,&0.972 &1100  & \underline{230}  &0.63 &14.2&\underline{7.2}&1.12&8.7&15.9&0.7&10.5    &2,3,7,8\\ 
1E\,0657-558&Bullet                    &0.296&1660  &170      &0.73 &12.4&\underline{12.3}&2.01&12.7&24.9&0.3&13.3&2,3,7,9\\
A\,1367&Leo                                &0.022&\underline{760}&210    &8.1 & 3.7 &1.1&\underline{0.11}&1.3&\underline{0.4}&0.5&1.2  &5,6\\ %11h44m29.5s +19o50'21''
A\,1656&Coma                            &0.023&1310&290        &10.5&6.9&2.9  &\underline{1.0}&3.5   &\underline{6.7}&\underline{0.5}&5.6   &6,10,11\\ %12h59m48.7s +27o58'50''
Virgo             &                &0.004&770        &    310          &62.5&   2.4& 2.7  &1.5 &8.0 &\underline{3.6}&0.3 &5.3  &12,13,14\\ %12h27m =+12o43'
A\,1991&                          &0.059  & 730 &60    &0.90   &2.3  &6.4 &0.1  &1.23 &\underline{0.2}& \underline{0.5}&3.5                &5,6,15\\ %14h54m30.2s +18o37'51''
ST\,J2106-5844&             &1.132  & 960 & 200   &0.54 &    9.4  &\underline{11.5}&1.17&7.1&11.1&0.3  &14.6          &2,3,7\\
ST\,J2248-4431&AS\,1063&0.348  &1630& 370  &1.4 &     11.5& \underline{2.9}&1.95&13.1&22.3&0.3  &7.0          &2,3,7\\
ST\,J2344-4243&Phoenix&0.596&1330&290       &0.88     &  14.9&\underline{4.8}&1.48 &9.6& 22.1  &0.5&8.8    &2,3,7,16\\  \hline
\multicolumn{14}{l}{}\\ [-1mm]
\multicolumn{14}{l}{\aa\ A -- Abell, ST -- SPT-CL, AT -- ACT-CL, P -- PLCK.}\\
\multicolumn{14}{l}{\bb\ The radius bounding the mass $M_{500}$
  of a cluster with a mean density equal to $500\ \rho_{\rm cr}(z)$ of the Universe.}\\
\multicolumn{14}{l}{\cc\ The gas density at the cluster center, $10^{-3}\ \mbox{\rm cm}^{-3}$.}\\
\multicolumn{14}{l}{\dd\ The microwave background scattering
  ``efficiency'' (Kravtsov et al. 2006), $10^{14}\ M_{\odot}\ \mbox{keV}$.}\\ 
\multicolumn{14}{l}{\ee\ The abundance of the iron-group
  elements compared to the normal cosmic abundance.}\\ 
\multicolumn{14}{l}{\ff\ The Thomson optical depth along the
  line of sight passing through the cluster center.}\\
\multicolumn{14}{l}{\g2\ 1 -- Menanteau et al. (2012); 2 --
  Bulbul et al. (2019);  3  --  Bleem et al. (2015); 4 -- White et al. (1997);}\\
\multicolumn{14}{l}{\ \ \  5 -- Jones and Forman (1984); 6 --
  David et al. (1993); 7 -- Williamson et al. (2011); 8 -- Aghanim et al. }\\ 
\multicolumn{14}{l}{\ \ \  (Planck
  Collaboration) (2011); 9  --  Markevitch et al. (2002); 10 --
  Ade et al. (Planck Collaboration) (2013);}\\ 
\multicolumn{14}{l}{\ \ \ 11 -- Herbig et al. (1995); 12 --
  Forman and Jones (1982); 13  -- Ade et al. (Planck Collaboration) (2016b);}\\
\multicolumn{14}{l}{\ \ \  14 -- Gavazzi et al. (2009); 15  --
  Vikhlinin et al. (2006); 16 -- McDonald et al. (2015).}\\ 
\multicolumn{14}{l}{$^*$ The underlined estimates were obtained
  using the formulas and dependences from Navarro et al. (1997)
  and }\\
\multicolumn{14}{l}{\ \ \ Kravtsov et al. (2006).}
\end{tabular}
%\end{center}
\end{table*}
%--------------------------------------------------------------------------------------------------

As $z$ increases, the spectrum of the thermal plasma radiation
in the cluster is also shifted to low energies and its intensity
decreases. This is indicated by the thick (blue) lines in
Fig.\,\ref{fig:z2}. This shift (and attenuation) allows the
effects of Compton scattering and photoabsorption of the
background in distant clusters to be investigated at lower
energies. The fundamental difference in the shape of the $z$
dependence of the amplitude of Compton X-ray background
distortions compared to the flux of bremsstrahlung and
recombination radiation is illustrated by Fig.\,\ref{fig:z}. The
solid (dark green) lines in this figure indicate the variation
in the spectral flux of thermal radiation wth $z$ for a cluster
with a uniform density distribution, $\tau_{\rm
  T}=1.2\times10^{-2}$, $R_{\rm c}=350$ kpc, and $kT_{\rm e}=2$
keV expected during its observation by a telescope with an
aperture radius of $5$\arcmin\ (FWHM). The curves from top to
bottom correspond to the fluxes at energies $h\nu=2,\ 5,\ 12$
and $20$ keV. The dashed line indicates the $z$ dependence of
the integrated flux. The initial flux rise up to $z\simeq0.06$
is related to the increase in the volume emission measure of the
cluster gas visible within the aperture:
$$ EM=\frac{4\pi}{3} R_{\rm c}^3\left[1-\left(1-\frac{\rho_{\rm
      z}^2}{R_{\rm c}^2}\right)^{3/2}\right],$$ where $\rho_{\rm
  z}$ is the impact parameter in the cluster frame corresponding
to the specified aperture width. At high $z$ z, when the cluster
is already completely within the field of view, the integrated
flux of its thermal radiation drops with $z$ as a power law; the
spectral fluxes at high energies drop more rapidly due to the
cutoff of the bremsstrahlung and recombination radiation
spectrum at the corresponding energies. The dotted line
indicates the X-ray background flux at 5 keV falling into this
aperture.

Because of the shift of the photoabsorption lines and the
absorption feature related to the recoil effect toward low
energies, the $z$ dependence of the amplitude of the Compton
distortions and photoabsorption takes a fairly complex shape
(Fig.\,\ref{fig:z}, especially $6$ and $8$ keV). The abrupt
jumps on these curves are associated with the passage of the
threshold of the absorption line at $\sim 9$ keV as $z$ changes,
given its fine structure. Such jumps should also be observed at
low energies --- leftward of the absorption threshold at $\sim2$
keV. We emphasize that this figure shows the amplitude of the
absolute background distortions (i.e., the difference in the
spectral flux of radiation within the aperture compared to the
initial spectrum taken in absolute value). The previous figures
presented the relative distortion amplitude (in \%). Here (just
as above when considering the thermal cluster radiation) we took
into account the effect of observation of only the part of the
cluster at low $z$ (due to the excess of its size above the
aperture size) and the effect of observation of a noticeable
fraction of the undistorted background at high $z$ (because the
cluster begins to occupy only part of the aperture). In clusters
with a smooth density distribution (described by the $\beta$ or
NFW models --- see the Appendix) the curves in
Fig.\,\ref{fig:z}, modified by these effects will be smoother.

As follows from the figure, the amplitude of the
background distortions due to scattering and absorption
changes with $z$ much more weakly than the flux
of thermal radiation. In this case, the probability of
detecting the distortions in the background spectrum
from the interaction with the gas of distant clusters
may turn out to be even higher than that from the
interaction with the gas of nearby clusters. In any
case, the detection of such distortions remains a very
challenging problem.

%xxxxxxxxxxxxxxxxxxxxxxxxxxxxxxxxxxxxxxxxxxxxxxxxxxxxxxxxxxxxxxx
\section*{INDIVIDUAL CLUSTERS}
\noindent
Table\,\ref{table:clusters} gives basic characteristics (the
temperature and central density of the intergalactic gas, other
parameters of the $\beta$ model density distribution) for
several known rich clusters that exhibit strong microwave
background radiation distortions. In particular, this can be
seen from the high values of the parameter $Y_{\rm SZ}$
characterizing the amplitude of the background distortions
(Kravtsov et al. 2006) given in column 11 of
Table\,\ref{table:clusters}. Such clusters as Phoenex, SPT-CL
J0615-5746, SPT-CL\,J2106-5844, and El Gordo were even
discovered owing to this effect --- by SPT (Williamson et
al. 2011), ACT (Menanteau et al. 2012), or the Planck satellite
(Aghanim et al.  (Planck Collaboration), 2011). These are mostly
very massive hot clusters with $kT_{\rm e}\ga10$ keV, but the
cold nearby Virgo, A\,1367, and A\,1991 clusters were also
included in the list. The detectability of distortions in the
X-ray background from a cluster, and this list was compiled
precisely for its estimation, is determined by many factors, and
a reliable detection of the effect in the microwave background
by no means implies that it can be detected in X-rays.
%--------------------------------------------------------------------------------------------------
\begin{figure}[t]
\hspace{-1mm}\includegraphics[width=0.99\linewidth]{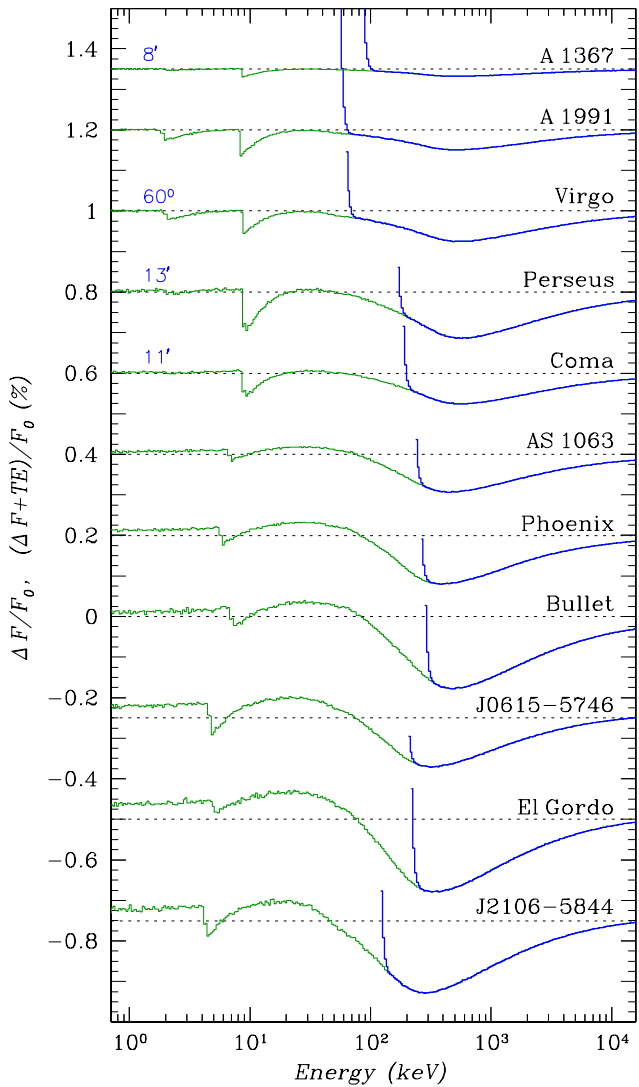}
%\epsffile{szxray_cl.ps}
%\fbox{\rule{0cm}{9cm}\rule{0.97\linewidth}{0cm}}

\caption{\rm Expected distortions in the spectrum of the X-ray
  background intensity due to Compton scattering and
  photoabsorption in the hot gas of several known galaxy
  clusters (thin green lines). The thick (blue) lines also take
  into account the thermal gas radiation.\label{fig:clusters}}
\vspace{-4.5mm}
\end{figure}
%--------------------------------------------------------------------------------------------------

For a number of clusters we failed to find the measured values
of some density distribution parameters.  These parameters were
then estimated from their dependence on $M_{500}$ ($Y_{\rm SZ}$)
determined by Vikhlinin et al. (2006) and Kravtsov et
al. (2006). In Table\,\ref{table:clusters} their values are
underlined for clarity. For all clusters we calculated the
optical depth for Thomson scattering along the line of sight
passing through their center (column 13 in the table). Note that
for such nearby and extended clusters as Virgo, Coma, and
Perseus it turns out to be only a factor of 2--3 smaller than
the optical depth of distant supermassive clusters like El
Gordo, Bullet, SPT-CL J2106-5844, and SPT-CL J0615-574. For
convenience, column 6 gives the angular sizes of the clusters
corresponding to the values of $R_{\rm c}$ specified in
Table\,\ref{table:clusters}. For all clusters in the table we
computed the relative distortions produced by them (the gas kept
by their gravity) in the X-ray background radiation spectrum by
the Monte Carlo method under the assumption of a $\beta$ density
distribution of the intergalactic gas. The computations were
carried out for the same exponent of the distribution
$\beta=2/3$, because the scatter of $\beta$ values for separate
clusters did not exceed the errors of their determination.  The
abundance of the iron-group elements $Z$ in the gas relative to
the cosmic abundance (column 12 in the table) was taken from the
literature. The cluster redshifts were taken into account.

The results of our computations of the relative background
radiation distortions (in percent to the initial spectrum) are
presented in Fig.\,\ref{fig:clusters}. They correspond to the
observations toward the cluster center by a telescope with a
narrow aperture corresponding to the cluster core radius
($\sim1$\arcmin\ for most clusters, see
Table\,\ref{table:clusters}). As one recedes from the cluster
center or when observing by a telescope with a wide aperture,
the distortion amplitude should drop. The thin (green) and thick
(blue) lines indicate the background distortions proper and the
distortions including the thermal gas radiation,
respectively. We see that the rich hot clusters in the lower
part of the figure lead to large deviations of the background
over the entire spectrum --- positive at $h\nu\la100$ keV (the
excess radiation due to the Doppler effect) and negative at
energies above $h\nu\ga100$ keV (the dip in the spectrum due to
the recoil effect). The feature at $\sim9$ keV associated with
photoabsorption by iron is noticeably weaker in amplitude than
the MeV dip --- the iron at temperatures typical for such
clusters is almost fully ionized. We also see that the intrinsic
thermal gas radiation presents a huge problem for the detection
of the effect being discussed at energies $\la 200$ keV.

For the cold clusters in the upper part of the figure, such as
Virgo, A\,1991, Coma, and Perseus, the amplitude of the positive
background distortions at low energies is negligible, while the
amplitude of the MeV dip remains fairly large; it differs from
the amplitude of the dip for the SPT-CL\,J2106-5844,
SPT-CL\,J0615-5746, and El Gordo clusters, which were believed
to be most massive in the Universe, only by a factor of 2 or
3. The features due to photoabsorption in the spectrum of the
background distortions by these clusters are expectedly larger
than those for the hot clusters due to the low gas temperature
--- the feature at 9 keV is even comparable in amplitude to the
dip at high energies. The thermal radiation from the clusters,
also expectedly, begins to hinder the detection of the effect at
appreciably lower ($h\nu\la50$ keV) energies than that for the
rich clusters. The figure suggests that for the Virgo and
A\,1991 clusters the drop in the background due to recoil effect
can be searched for already at energies 60--100 keV without any
noise related to the thermal gas radiation. It is important that
the Virgo cluster, along with Perseus, Coma, and A\,1367 have
noticeable angular sizes (see the left part of
Fig.\,\ref{fig:clusters}, where the angular core radius of these
clusters is presented in arcmin/deg).  {\it The observations of
  such extended clusters even at a slightly smaller amplitude of
  the background distortions than that for rich, but distant
  clusters may turn out to be much more significant and
  fruitful.\/}

%xxxxxxxxxxxxxxxxxxxxxxxxxxxxxxxxxxxxxxxxxxxxxxxxxxxxxxxxxxxxxxx
\section*{X-RAY BACKGROUND FLUCTUATIONS}
\noindent
The detection of hard X-ray background deviations at a level of
fractions of percent is not something absolutely unattainable
per se. Such measurements aimed at searching for background
fluctuations have already been carried out by both HEAO-1 (Boldt
1987; Treyer et al. 1998) and RXTE (Gruber et al. 1999b;
MacDonald et al. 2001) observatories.

In particular, the
HEXTE/RXTE instrument performed almost simultaneous observations
of two sky regions spaced $3^{\circ}$ apart. The differences in the
fluxes measured in these regions within the field of view of the
instrument with an area of 1.1 sq. deg allowed background
fluctuations much weaker than was possible in individual
observations to be searched for. Such fluctuations were actually
detected at a flux level of $(0.092\pm0.014),$ $(0.11\pm0.02),$
and $(0.23\pm0.08)$\% of the background spectrum in the energy
ranges 15--20, 20--25, and 34--41 keV, respectively. The
relative amplitude of the fluctuations is seen to have increased
with energy. Similar results, in both background fluctuation
level and their energy dependence, were previously obtained by
the A-2 instrument of the HEAO-1 observatory.

The authors explained their results by a nonuniform distribution
of matter in the Universe and, accordingly, by a nonuniform
distribution of AGNs, which are to a large extent responsible
for the hard X-ray background. The energy dependence of the
fluctuation level can then be explained by the drop in the
effective number of sources responsible for the background and
fluctuations when passing to higher energies. However, it may
well be that not only AGNs, but also the distortions forming in the
spectrum of the background radiation as it passes through the
hot intergalactic gas in unresolved distant clusters of galaxies were
one of the causes of the detected background fluctuations. At
least the relative amplitudes of the detected fluctuations agree
well with the results of our computations presented in this
paper.

Therefore, the conclusion reached by Boughn and Crittenden
(2004, 2005) about a significant correlation of the fluctuations
in the distribution of the X-ray background measured by the
HEAO-1 observatory and the microwave background measured by the
WMAP satellite is also of interest. It seem quite probable that
in both energy ranges we are dealing with the fluctuations associated
with the same phenomenon --- the cosmic background scattering in
the hot gas of clusters of galaxies. Estimates and detailed
computations of the role of distant clusters in the formation of
the microwave background fluctuations have been performed at
some time by Longair and Sunyaev (1969) and Markevitch et
al. (1992).

The fact of achieving such a high sensitivity when observing the
X-ray background fluctuations in wide sky fields confirms that
the study is important and promising for detecting the effect
being discussed in nearby extended galaxy clusters, such as
Virgo, Coma, A\,426, and A\,1367.

%xxxxxxxxxxxxxxxxxxxxxxxxxxxxxxxxxxxxxxxxxxxxxxxxxxxxxxxxxxxxxxx
\section*{SCATTERING OF RADIATION FROM ACTIVE GALAXIES OF A CLUSTER}
\noindent
Some of the galaxy clusters may have active nuclei (AGNs), whose
hard radiation should be subject to scattering and absorption in
its hot gas, like the background radiation. The scattered
diffuse X-ray radiation from such an AGN can be perceived as an
additional background distortion.

\begin{figure}[t]
  \vspace{-1.5mm}
  \hspace{-1mm}\includegraphics[width=1.01\linewidth]{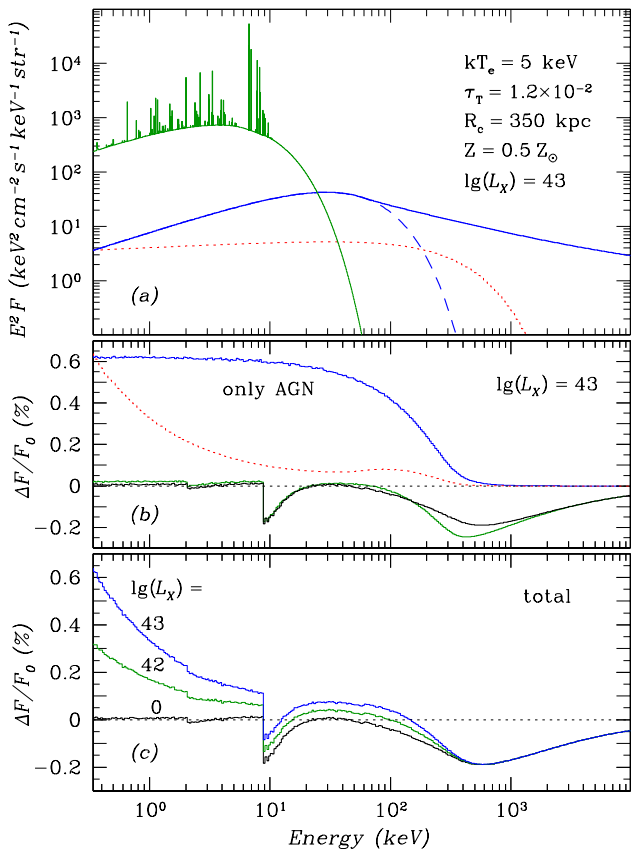}
%\epsffile{szxray_agn.ps}
\caption{\rm Influence of the AGN radiation on the observations
  of background distortions in the hot cluster gas: (a) the AGN
  radiation spectrum (red dotted line) in comparison with the
  background and cluster gas spectra; (b) the distortions of the
  AGN (green) and background (black curves) spectra, the
  scattered AGN radiation relative to its initial spectrum (blue
  curve) and relative to the initial background spectrum (red
  dotted line); (c) the background distortions including the
  scattered AGN radiation. The optical depth of the gas toward
  the center is $\tau_{\rm T}=1.2\times10^{-2}$, the cluster
  radius is $R_{\rm c}= 350$ kpc, $kT_{\rm e}=5$ keV, and
  $Z=0.5\ Z_{\odot}$. The AGN X-ray luminosity is $L_{\rm X}=
  1\times 10^{43}\ \mbox{erg s}^{-1}$, the green line on the
  lower panel indicates the case of $L_{\rm X}= 1\times
  10^{42}\ \mbox{erg s}^{-1}$.
  \label{fig:agn}}
\vspace{-5mm}
\end{figure}

Can this radiation hinder the detection of the effect being
discussed in the paper? Let us place an AGN with an X-ray
luminosity $L_{\rm X}=1\times10^{43}\ \mbox{erg s}^{-1}$ and a
power-law spectrum with an exponential cutoff at high energies
typical for AGNs at the center of our standard model cluster
with a uniform density, optical depth $\tau_{\rm
  T}=1.2\times10^{-2}$, core radius $R_{\rm c}=350$ kpc,
electron temperature $kT_{\rm e}=5$ keV, and metallicity
$Z=0.5\ Z_{\odot}$. We will take the photon index of the
spectrum to be $\alpha=1.9$ and the cutoff energy to be $E_{\rm
  c}=300$ keV (Sazonov et al. 2008; Ueda et al. 2014). The
radiation spectrum of such an AGN is indicated in
Fig.\,\ref{fig:agn}a by the dotted red line for comparison with
the spectra of the hot gas in the galaxy cluster and the cosmic
background radiation.  Note that the X-ray luminosity of the
thermal radiation from the cluster gas is $L_{\rm c}=6.9 \times
10^{44}\ \mbox{erg s}^{-1}$. The green line at the bottom in
Fig.\,\ref{fig:agn}b indicates the spectrum of the relative
distortions arising in the AGN radiation when it passes through
the hot cluster gas. It is very similar to the spectrum of the
relative distortions arising in the background radiation
(indicated in the same figure by the black line). Note only the
shape of the MeV dip slightly skewed to low energies, which,
obviously, reflects the difference in shape between the initial
AGN radiation spectrum and the background spectrum (primarily
the exponential cutoff at high energies in the AGN spectrum).
%--------------------------------------------------------------------------------------------------
%--------------------------------------------------------------------------------------------------

In reality, however, we are interested not in the distortions in
the AGN radiation spectrum, but in its radiation that was
scattered in the cluster and became diffuse, because the direct
escape radiation will be perceived as the radiation of a compact
source (AGN) when analyzing the data and, naturally, should be
subtracted.

The blue line in Fig.\,\ref{fig:agn}b indicates the spectrum of
scattered AGN photons (relative to its initial spectrum), while
the red dotted line indicates the same spectrum, but relative to
the initial background radiation spectrum. Remarkably, the
scattered photon spectrum is smooth and does not contain any
negative features related to photoabsorption or the recoil
effect after scattering. Its amplitude relative to the initial
AGN radiation spectrum is more than 0.6\%, i.e., it is much
greater than the amplitude of the final AGN radiation
distortions. We noted these properties of the scattered
radiation (only that of the background) previously when
discussing Fig.\,\ref{fig:1scat}. The amplitude of the scattered
AGN radiation relative to the background spectrum turns out to
be small everywhere, except the standard X-ray band $h\nu\la10$
keV. The enhancement of the distortions in the soft X-ray band
is related to the differences in the AGN and background spectra:
the AGN intensity here begins to approach and even exceed the
background intensity. However, many AGNs exhibit a low-energy
cutoff in the radiation spectrum related to the absorption of
their radiation in the gas-dust torus surrounding the
supermassive black hole in the galactic nucleus. Clearly, when
the absorption is taken into account, the rise of the relative
background distortions due to the soft X-ray AGN radiation
should be less distinct or even disappear altogether.

The important thing is that the spectrum of the scattered AGN
radiation contains no absorption features and, therefore, it
cannot reduce or smear such features in the spectrum of the
background distortions arising when it passes through the hot
cluster gas. In the case of very bright AGN flares occurring on
a time scale of hundreds of thousands of years, narrow lines
associated with resonance scattering of the AGN X-ray emission
by Fe, S, and Si ions can appear in the background spectrum at
these energies (Sazonov et al. 2002).

Indeed, as Fig.\,\ref{fig:agn}c shows, adding the AGN radiation
scattered in the cluster gas to the spectrum of the emerged
background distortions does not smear the features related to
background photoabsorption on the $K$ and $L$ shells of strongly
ionized iron and nickel ions in the cluster gas. What is
especially remarkable, the scattered AGN radiation does not
reduce and does not smear the MeV dip in the background spectrum
related to the recoil effect when its photons are scattered by
electrons in the cluster gas.

At the same time, the AGN radiation can lead to a rise in
background intensity at low energies $h\nu\la9$ keV and in the
hard X-ray range 20--150 keV, making it difficult to correctly
interpret the background distortions at these energies. What is
especially bad, the scattered AGN radiation is subject to the
X-ray echo effect (Fabian 1977; Vainshtein and Sunyaev 1980),
which leads to the conservation of the scattered X-ray radiation
from the AGN for many ($\Delta t\sim R_{\rm c}/c\simeq1.1$ Myr,
where $c$ is the speed of light) years after the decay of the
X-ray activity of the AGN itself. Similar radiation of the
former activity of the black hole at the center of our Galaxy
was observed from molecular clouds (Markevitch et al. 1993;
Sunyaev et al. 1993). On the other hand, at such a long time of
the reaction to AGN activity ($\simeq1$ Myr), it will also take
a long time for the scattered radiation to be completely formed
and to reach a stable intensity level. Thus, the influence of
the AGN radiation can be effectively below our estimates.

Obviously, our analysis is completely extended to the case where
the AGN is located not inside the cluster, but at some distance
behind it or even near it; it is only necessary to take into
account the area of the solid angle at which the cluster is seen
from the AGN location.

%xxxxxxxxxxxxxxxxxxxxxxxxxxxxxxxxxxxxxxxxxxxxxxxxxxxxxxxxxxxxxxx
\section*{BACKGROUND DISTORTIONS IN A WARM INTERGALACTIC MEDIUM}
\noindent
At $z\la 1$ the total mass of the visible matter contained in
the stars, dust, and hot gas of galaxy clusters and the clouds
of atomic and molecular hydrogen accounts for only $\sim 1/3$ of
the entire baryonic mass of the Universe (Fukugita et
al. 1998). At the same time, observations of the ``forest'' of
$Ly_{\alpha}$ hydrogen lines at $z\sim3$ give a factor of
$\sim3$ greater value consistent with the mean baryon density in
the Universe $\Omega_{\rm B}=0.044$ (Rauch 1998). It is believed
that $\sim 2/3$ of the baryons that have eluded observation are
contained in the moderately hot plasma with a temperature $\sim
10^6$ K located in filaments and other similar structures on the
far periphery of galaxy clusters (Cen and Ostriker 1999). This
gas phase is called the Warm-Hot Intergalactic Medium
(WHIM). The existence of WHIM was confirmed by the detection of
a soft X-ray excess from a plasma with $kT\sim0.2$ keV with a
prominent blend of the O\,VII triplet at 0.57 keV and the
O\,VIII $Ly_{\alpha}$ line at 0.65 keV in the spectra of several
galaxy clusters (see, e.g., Finoguenov et al. 2003; Kaastra
2004a, 2004b).
%--------------------------------------------------------------------------------------------------
\begin{figure}[t]
\hspace{-2mm}\includegraphics[width=1.02\linewidth]{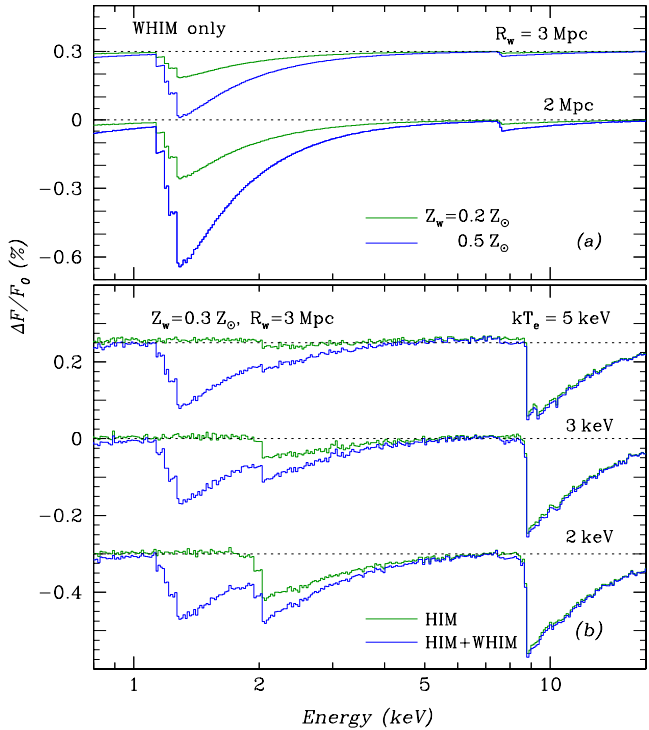}
%\epsffile{szxray_whim.ps}

\caption{\rm Expected distortions in the X-ray background
  spectrum due to scattering and absorption in the envelope of
  the warm-hot intergalactic medium (WHIM) surrounding the
  galaxy cluster toward its center: (a) without the background
  distortions in the hot cluster gas and (b) with the
  distortions (the green lines indicate the background spectra
  before the distortion in WHIM). The hot gas (HIM) in the
  cluster has a Thomson optical depth toward the center
  $\tau_{\rm T}=1.2\times10^{-2}$, core radius $R_{\rm c}= 350$
  kpc, electron temperature $kT_{\rm e}=2,\ 3,\ 5$ keV, and
  metallicity $Z=0.5\ Z_{\odot}$; the WHIM envelope has a factor
  of 4 larger mass, radius $R_{\rm w}= 2\ \mbox{\rm or}\ 3$ Mpc,
  electron temperature $kT_{\rm w}=0.2$ keV, and metallicity
  $Z_{\rm w}=0.2,\ 0.3,\ \mbox{or}\ 0.5\ Z_{\odot}$.
\label{fig:whim}}
\end{figure}
%--------------------------------------------------------------------------------------------------

Can the interaction of the X-ray background with WHIM give rise
to additional distortions in its spectrum? Suppose, for
simplicity, that each galaxy cluster is surrounded by a thick
spherical WHIM layer (shell) with a quasi-uniform density
distribution $\rho_{\rm w}=N_{\rm w}m_{\rm p}$. The outer radius
of the layer is $R_{\rm w}\simeq 2-3$ Mpc (Kaastra
2004a). Assuming that the hot gas in the cluster (with a mass
$M_{\rm g}$ within the radius $R_{\rm c}$) contains $\sim50$\%
of the visible matter, we find that the WHIM mass in the outer
layer is $M_{\rm w}\simeq (2/3\,\Omega_{\rm B})/
(0.5\times1/3\,\Omega_{\rm B}) M_{\rm g}\simeq 4\,M_{\rm g}.$
The mean electron density in this layer is then $N_{\rm w}=4
N_{\rm e} (R_{\rm c}/R_{\rm w})^3\simeq5.3\times10^{-5}\times$
$$\left(\frac{\tau_{\rm T}}{1.2\times10^{-2}}\right)
\left(\frac{R_{\rm c}}{350\ \mbox{\rm kpc}}\right)^{2}
\left(\frac{R_{\rm w}}{3\ \mbox{\rm Mpc}}\right)^{-3}\mbox{\rm
  cm}^{-3}.$$ This estimate agrees well with the WHIM density
measurements from the O\,VII and O\,VIII lines in the X-ray
spectra of galaxy clusters (Kaastra 2004a).  The electron
density in the hot cluster gas $N_{\rm e}$ expressed here via
its Thomson optical depth along the line of sight passing
through the cluster center, $\tau_{\rm T}=2 \sigma_{\rm T}N_{\rm
  e}R_{\rm c}=1.2\times 10^{-2}.$ The Thomson optical depth of
the WHIM envelope along the same line of sight is $\tau_{\rm
  w}\simeq 2\sigma_{\rm T}N_{\rm w}(R_{\rm w}-R_{\rm c})\simeq 4
\tau_{\rm T}(R_{\rm c}/R_{\rm w})^2\simeq
6.5\times10^{-4}\times$
$$ \left(\frac{\tau_{\rm T}}{1.2\times10^{-2}}\right)
\left(\frac{R_{\rm c}}{350\ \mbox{\rm kpc}}\right)^{2}
\left(\frac{R_{\rm w}}{3\ \mbox{\rm
    Mpc}}\right)^{-2}\!\!\!\!\!.$$ Obviously, the distortions
related to purely Compton scattering in WHIM will be the same as
those after scattering in the hot cluster gas (Hot
Integrgalactic Medium or HIM), but smaller in absolute value,
because the optical depth of the envelope is small. In this
case, the iron and nickel atoms in WHIM will be ionized much
more weakly than those in the hot cluster gas (at $kT_{\rm
  w}\simeq 0.2$ keV --- to Fe\,XIV--Fe\,XVII and
Ni\,XIII--Ni\,XVII, respectively).  Therefore, the
photoabsorption of the background radiation in the envelope
should be more efficient. Unfortunately, WHIM is most likely
much poorer in metals than even HIM (see, e.g., Finoguenov et
al. 2003); therefore, it is difficult to estimate this
efficiency without numerical computations.

Figure\,\ref{fig:whim} presents the results of such computations
of the distortions in the X-ray background spectrum toward the
cluster center arising when it passes through the WHIM envelope
surrounding the cluster. The distortions only in WHIM are shown
at the top (in Fig.\,\ref{fig:whim}a); the changes in the
spectrum that occurred in the hot cluster gas were disregarded
here. It will be possible to see the shown distortions in
cluster observations at large impact parameters $\rho\gg R_{\rm
  c}$. We considered the envelopes with various outer radii
$R_{\rm c}$ (and, accordingly, various densities $N_{\rm w}$ and
optical depths $\tau_{\rm w}$) and metallicities $Z_{\rm
  w}$. The WHIM temperature was taken to be $kT_{\rm e}=0.2$
keV. The absorption threshold on the $K$ shell of iron and
nickel is seen to be shifted leftward along the energy axis
compared to the spectrum of the distortions in HIM (from
$\simeq8.8$ to $\simeq7.6$ keV); the absorption threshold on the
$L$ shell is shifted even more strongly --- from $\simeq 2.03$
to $\simeq 1.27$ keV. Whereas the hard line has a negligible
depth, the depth of the absorption line at 1.27 keV is
unexpectedly large even for the lowest metallicity considered
$Z_{\rm w}=0.2\ Z _{\odot}$.

Absorption lines are also formed at energies below $\sim1$ keV,
but they are impossible to observe due to the intrinsic thermal
radiation from WHIM. Most of the lines of resonant scattering of
the X-ray background in WHIM predicted by Churazov et al. (2001)
also fall into this region. Above $\sim1$ keV there are only
narrow recombination emission lines at energies
$1.07,\ 1.13,\ 1.33-1.35,\ \mbox{and}\ 1.87$ keV in the WHIM
radiation spectrum. Note that iron and nickel ions with
vacancies on the lower electronic shells are formed as a result
of background photoabsorption in the WHIM plasma, which could
not be formed in it due to collisional processes. Although these
ions recombine mainly through the Auger effect, the formation of
photons of fluorescent lines (similar to the $K_{\alpha}$ line
with energy 6.4 keV emitted with a 34\% probability when neutral
iron atoms are ionized) is also possible. There are no such
emission lines in the intrinsic thermal recombination spectrum
of WHIM.  These lines are narrow and much harder to detect than
the photoabsorption lines; therefore, they are disregarded in
this paper.
%--------------------------------------------------------------------------------------------------
\begin{figure}[t]
 \hspace{-2mm}\includegraphics[width=1.02\linewidth]{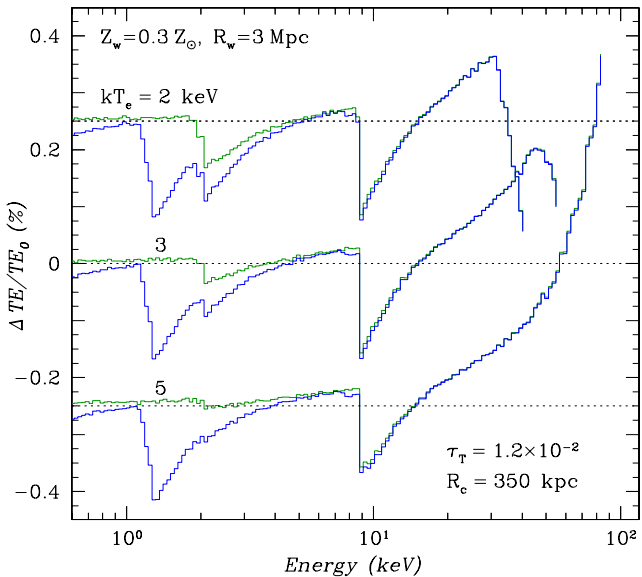}
%\epsffile{szxray_whimtbr.ps}

\caption{\rm Expected distortions in the thermal radiation
  spectrum of the hot cluster gas (HIM) in a galaxy cluster due
  to scattering and absorption in the envelope of the warm-hot
  intergalactic medium (WHIM) surrounding the galaxy cluster
  toward its center (blue lines). The hot gas (HIM) in the
  cluster has a Thomson optical depth toward the center
  $\tau_{\rm T}=1.2\times10^{-2}$, radius $R_{\rm c}= 350$
  kpc, electron temperature $kT_{\rm e}=2,\ 3,\ 5$ keV, and
  metallicity $Z=0.5\ Z_{\odot}$; the WHIM envelope has a factor
  of 4 larger mass, radius $R_{\rm w}= 3$ Mpc, optical depth
  $\tau_{\rm w}\simeq 6.5\times10^{-4}$, electron temperature
  $kT_{\rm w}=0.2$ keV, and metallicity $Z_{\rm
    w}=0.3\ Z_{\odot}$. The green lines indicate the cluster HIM
  spectra before the interaction with WHIM.
\label{fig:whimtbr}}
\end{figure}
%--------------------------------------------------------------------------------------------------

The total spectra including the distortions in both WHIM and HIM
are presented at the bottom (in Fig.\,\ref{fig:whim}b). The
radius of the WHIM envelope was taken to be $R_{\rm c}=3$ Mpc,
the temperature is again $kT_{\rm w}= 0.2$ keV, and the
metallicity is $Z_{\rm w}=0.3\ Z_{\odot}$. The optical depth of
the hot cluster gas is $\tau_{\rm T}=1.2\times 10^{-2}$, the
temperature is $kT_{\rm e}= 2,\ 3,\ \mbox{\rm or}\ 5$ keV, and
its metallicity is $Z=0.5\ Z_{\odot}$. The spectra of the
background distortions formed in HIM were previously presented
in Fig.\,\ref{fig:res.lines}. Substantial changes are seen to
occur in the background spectrum when passing through the WHIM
envelope: (1) a new intense broad absorption line appears at
$\sim1.3$ keV, which does not depend on the cluster hot gas
temperature; (2) the depth (equivalent width) of the absorption
line at $\sim2$ keV increases noticeably; and (3) the absorption
threshold at $\sim8.8$ keV is smeared.

Figure\,\ref{fig:whimtbr} shows how the effect being discussed
affects the spectrum of the intrinsic thermal (to be more
precise, bremsstrahlung) plasma radiation. Our computations were
done for the WHIM envelope and the hot cluster gas with the same
parameters as those in Fig.\,\ref{fig:whim}. The changes in the
thermal spectra related to the interaction directly with the
cluster gas (HIM) have already been shown in
Fig.\,\ref{fig:gas.edge}. Here they are represented by the thin
green lines. A broad absorption line with the threshold at
$\sim1.3$ keV and an amplitude of $\sim0.16-0.22$\% (relative to
the initial undistorted thermal spectrum) additionally appears
in these spectra after the interaction with WHIM.  Furthermore,
the line at $\sim2$ keV already present in the spectra of cold
clusters is enhanced. The profile and amplitude of the
absorption line at $\sim9$ keV barely change. Thus, for
favorable WHIM parameters two absorption lines at once, at $\sim
1.3$ and $9$ keV, should be present in the thermal radiation
spectra of the hot cluster gas.

The model used may be oversimplified and contains a number of
poorly known parameters (primarily those related to the
nonuniformity of the WHIM distribution). At the same time, it
clearly shows that the interaction with WHIM can introduce
noticeable distortions both into the cosmic X-ray background
spectrum and the thermal radiation spectrum of the cluster,
distortions comparable to those from their scattering and
absorption in the hot cluster gas. This effect cannot not be
neglected for nearby ($z\la1$) clusters.  The formation of the
absorption line at $\sim1.3$ keV, whose amplitude does not
depend on the hot gas parameters, but only on the WHIM
parameters, may turn out to be very important in light of
studying the properties of this mysterious matter itself in the
distant environment of clusters of galaxies.

%xxxxxxxxxxxxxxxxxxxxxxxxxxxxxxxxxxxxxxxxxxxxxxxxxxxxxxxxxxxxxxx
\section*{MAIN RESULTS}
\noindent
We presented the results of our numerical (Monte Carlo) and
analytical (solving the Kompaneets equation) computations of the
distortions arising in the spectrum of the X-ray and soft
gamma-ray background radiation as it passes through the hot
intergalactic gas in galaxy clusters. We investigated the
dependence of the distortion amplitude and shape on the
parameters of the cluster gas --- its temperature, optical
depth, and density distribution. The analogous distortions
arising in the microwave background radiation are well known, are
actively studied, and are widely used in observational
cosmology. We showed the following:

\begin{enumerate}
\item Compton scattering by electrons of the intergalactic gas
  in clusters leads to peculiar distortions of the background
  radiation --- a rise in its brightness at $h\nu\la 60$--$100$
  keV due to the Doppler effect and its drop at higher energies
  due to the recoil effect; the detection of background
  distortions allows the most important parameters of clusters
  and cosmological parameters to be measured;

\item the rise in background brightness in the hard X-ray range
  is proportional to the Compton parameter $y_{\rm C}=\tau_{\rm
    T}\ kT_{\rm e}/m_{\rm e}c^2,$ averaged over the visible
  (within the telescope aperture) part of the cluster (or the
  parameter $Y_{\rm SZ}$ in the case of a distant cluster), for
  the hottest clusters with $kT_{\rm e}\sim15$ keV it reaches
  $\sim 0.1$\% at energies $\sim100$ keV;

\item the decrease in background brightness due to the recoil
  effect has a maximum at energies $\sim500-600$ keV (in the
  cluster rest frame) and is proportional to the gas surface
  density or its Thomson optical depth $\tau_{\rm T}$ averaged
  over the visible (in the aperture) part of the cluster, it has
  an amplitude of $\sim0.2$--$0.3$\% for the optically thickest
  clusters and does not depend on the temperature;

\item photoabsorption by strongly ionized ions of the iron-group
  elements also leads to a decrease in the background with the
  formation of two absorption lines of characteristic shape with
  threshold energies $h\nu\sim 2$ and $9$ keV (in the cluster
  rest frame) in its spectrum, the amplitude of the features
  drops with rising temperature;

\item the interaction of the background with the colder
  ($kT\sim10^6$ K) plasma located on the distant ($\la 3$ Mpc)
  periphery of nearby ($z\la1$) galaxy clusters noticeably
  enhances the absorption line at $h\nu\sim 2$ keV in its
  spectrum and, moreover, leads to its splitting (the appearance
  of a broad intense satellite line at $h\nu\sim 1.3$ keV whose
  amplitude does not depend on the properties of the
  high-temperature cluster gas); the line detection will allow
  the parameters and distribution of this mysterious medium
  containing 2/3 of the baryonic matter in the Universe to be
  investigated;

\item features due to absorption and scattering are also formed
  in the intrinsic thermal and recombination radiation spectrum
  of the hot cluster gas; when the thermal spectrum of an
  optically thin plasma is subtracted from the observed total
  spectrum, they are perceived as additional background
  distortions with an amplitude $\sim100$\% or more;

\item the spectral shape of the background distortions depends
  on the cluster redshift $z$ (in contrast to the spectral shape
  of the microwave background distortions), although the
  distortion amplitude does not depend on $z$;

\item the detection of background distortions at energies
  $h\nu\la 20\ kT_{\rm e}$ is complicated by the presence of
  intrinsic thermal radiation from the intergalactic gas, which
  makes it very difficult to measure the effect of a rise in the
  background in the X-ray range; the observations of the MeV dip
  in the background spectrum are free from this noise for most
  clusters;

\item the detection of X-ray background distortions is also
  complicated by the presence of AGNs in (or near) the cluster
  with luminosities $L_{\rm X}\sim10^{41}-10^{43}\ \mbox{erg
    s}^{-1}$ at present or even in the distant (up to $\sim1$
  Myr due to the X-ray echo) past, because the X-ray radiation
  from such galaxies scattered in the cluster becomes diffuse
  radiation that enhances the background at $h\nu\la 150$ keV;
  the scattered radiation does not contain any photoabsorption
  lines and even the MeV dip in its spectrum and, therefore,
  does not change their amplitude in the spectrum of the
  background distortions.
\end{enumerate}

We considered a number of real galaxy clusters
and predicted the shape and amplitude of the spectra
of the background deviations expected for them
based on the available parameters. Although the
background intensity distortions toward the cluster
center are maximal for the most massive hot clusters,
cold nearby clusters like Virgo and Coma have real
chances for detecting the effect in the hard X-ray and
gamma-ray range already now. It seems possible that
the hard X-ray background fluctuations detected by
the HEAO-1 and RXTE observatories are associated
in part with the effect being discussed in the paper ---
with the background distortions in distant galaxy
clusters.

%xxxxxxxxxxxxxxxxxxxxxxxxxxxxxxxxxxxxxxxxxxxxxxxxxxxxxxxxxxxxxxx
\section*{CONCLUSIONS}
\noindent
It is clear from the foregoing that the main obstacle to
observing the distortions in the cosmic background spectrum due
to scattering by electrons in the X-ray range $h\nu\la 60-100$
keV (in the cluster rest frame) is the intrinsic plasma thermal
radiation\footnote{Besides, although the background distortions
  due to scattering and photoabsorption grow with increasing
  $\tau_{\rm T}$ of the cluster gas, the thermal radiation
  intensity, proportional to $N_{\rm e}^2$, grows much
  faster.}. If it were not for the illumination by this
radiation, modern X-ray telescopes with grazing-incidence optics
like NuSTAR (Harrison et al. 2013) might well observe these
distortions (or at least they have fallen just short of
this). The X-ray calorimeters with the resolution and
sensitivity required to measure the absorption line profile in
the background being designed at present would be able to
determine the gas temperature in a cluster and its composition
only from the absorption line profile in the background.

In December 2019 the SRG orbital astrophysical observatory with
the highly sensitive ART-XC and eROSITA telescopes onboard is
going to begin to scan the sky in X-rays for four years. The
ART-XC telescope, operating in the range 5--30 keV (Pavlinsky et
al. 2018), will be able to observe the distortions in the
background spectrum due to scattering by electrons in the hot
gas of clusters and photoabsorption by iron and nickel ions; the
eROSITA telescope, operating in the range 0.3--8 keV (Predehl et
al. 2018), will be able to investigate the absorption lines in
the background spectrum at lower energies. Even before the
survey, within the Russian part of the Performance Verification
Program, the SRG observatory is going to scan the nearby bright
Coma cluster.
 
We could attempt to subtract the thermal radiation spectrum of
the cluster gas itself from the measurements using theoretical
models and knowing the gas temperature, density, and
metallicity. However, (1) in real clusters the gas temperature
usually changes with radius and in a rather complicated way,
which greatly complicates highly accurate simulations; (2)
because of the finite optical depth of the cluster gas, the
distortions due to scattering and absorption should appear in
the spectrum of its thermal radiation, which are much greater in
absolute value than the background distortions that are added to
them when subtracting the ideal spectrum of an optically thin
plasma; (3) the intense thermal radiation of the gas should
introduce a noticeable statistical error into the background
measurements, reducing the signal-to-noise ratio $S/N$.
Therefore, to detect the effect, it is more preferable to
investigate cold, massive, nearby (extended) clusters using the
peripheral cluster regions, along with the central ones, for
observations.

The presence of AGNs in a galaxy cluster can additionally
complicate the X-ray observations of the effect, because their
radiation scattered in the cluster gas will be perceived as a
``positive'' background distortion. Scattered radiation can
exist in a cluster even hundreds of thousands of years after the
decay of the galactic nucleus --- due to the ``X-ray echo''
effect.  At the same time, it contains no absorption lines and
MeV dip in the spectrum and does not distort these features
appearing in the X-ray background spectrum when interacting with
the hot cluster gas.
  
It may well be that it would be more preferable to search for
the Compton cosmic background distortions in the hard X-ray and
gamma-ray spectral ranges, where their relative amplitude is
maximal and the contribution of the thermal radiation from the
hot cluster gas falls exponentially. However, it should be borne
in mind that the absolute amplitude of the distortions in these
ranges decreases due to the peculiar shape of the X-ray
background spectrum (see Fig.\,\ref{fig:z} above). Besides, the
instruments operating in these ranges are still noticeably
inferior in their capabilities to modern X-ray telescopes,
despite the abundance of problems and the general understanding
of the importance of studies in this field. The incessant
efforts (see, e.g., Fryer et al. 2019) to design more sensitive
MeV gamma-ray telescopes like e-ASTROGAM (Tatischeff et
al. 2016; De Angelis et al. 2017) or ASTENA/LAUE (Virgilli et
al. 2017) allow one to look with optimism at the prospects for
detecting the effect in this range in the immediate future.

Concluding this paper, we emphasize once again that all of the
expected background distortions in the hot cluster gas have a
very small ($\la1$\%, and as a rule, $\sim0.1$\%) absolute
value. They cannot be seen visually just by examining the
measured cluster spectrum. They will manifest themselves only in
the relative distortion spectrum, at a very high accuracy of
measurements and in proper simulations of the background spectra
and the thermal spectrum of the cluster gas, when all of the
effects discussed in the paper are taken into account.

\vspace{6mm}

%***************************************************************
\section*{ACKNOWLEDGMENTS}
\noindent
S.\,A. Grebenev is grateful to R.\,A. Burenin and
S.\,Yu. Sazonov for the useful discussion and consultations and
the Basic Research Program 12 of the Russian Academy of Sciences
(``Questions of the Origin and Evolution of the Universe with
the Application of Methods of Ground-Based Observations and
Space Research'') --- for financial support.  R.\,A. Sunyaev is
grateful to the Russian Science Foundation for support by grant
no. 19-12-00369.  This study was performed within the
``Universe'' theme of the research program at the Space Research
Institute of the Russian Academy of Sciences.
%****************************************************************
%---------------------------------------------------------------------------------------------
\begin{figure*}[t]
\centerline{\includegraphics[width=0.74\textwidth]{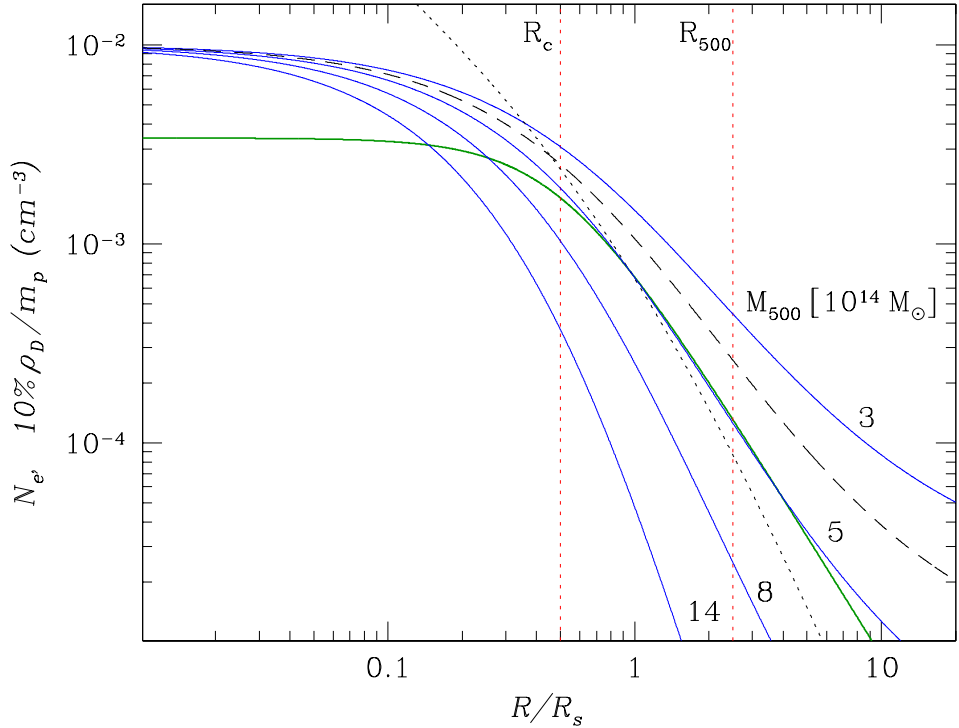}\hspace{5mm}}
%\epsfxsize=0.74\textwidth
%\epsffile{nfw.ps}
%\fbox{\rule{0cm}{6cm}\rule{0.97\linewidth}{0cm}}

\caption{\rm Equilibrium density distributions of an isothermal
  gas in the gravitational field of a cluster with dark matter
  having the NFW profile (Navarro et al. 1997) and their
  dependence on cluster mass $M_{500}$. The thin solid (blue)
  lines indicate the distributions with $kT_{\rm e}=5$ keV. The
  central gas density is everywhere assumed to be the same (see
  the text). The dashed (black) line for $M_{500}=5\times
  10^{14}\ M_{\odot}$ indicates the distribution with $kT_{\rm
    e}=6$ keV. The concentration parameter is $c=2.5$. The thick
  (green) line indicates the $\beta$ gas distribution with
  $\beta=2/3$ and $R_{\rm c}=0.5\, R_{\rm s}$ (the normalization
  ensures that the $\beta$ distribution near $R\sim R_{\rm s}$
  coincides with the gas distribution in the NFW model for
  $M_{500}=5\times 10^{14}\ M_{\odot}$; the dotted line
  indicates the 10\%\ level of the dark matter density
  distribution $\rho_{\rm D}/m_{\rm p}$ for this case).
\label{fig:nfw.prof}}
\end{figure*}
%---------------------------------------------------------------------------------------------
\vspace{1cm}

\begin{appendix}
%===========================================
\section{THE NAVARRO--FRENK--WHITE MODEL}
\noindent
The dark matter density distribution in galaxy clusters is
successfully described by the Navarro-Frenk-White (hereafter
NFW) profile (Navarro et al.  1997) found by N-body simulations
of equilibrium configurations in the theory of hierarchical
clustering of cold dark matter:
\begin{equation}\nonumber
 \rho_{\rm D} (R)=\rho_{\rm s}\frac{R_{\rm s}}{R}
  \left(1+\frac{R}{R_{\rm s}}\right)^{-2}.
\end{equation}
Here, $\rho_{\rm s}$ is the dark matter density parameter and
$R_{\rm s}$ is the scale parameter (cluster core radius). For
such a density profile the cluster dark matter mass within
radius $R$ is
\begin{equation}\label{eq:nfw.md}
M_{\rm D} (\!<\!R)=4\pi\rho_{\rm s}\,R_{\rm s}^3
 \left[\ln\left(\frac{R}{R_{\rm s}}+1\right)-\frac{R}{R+R_{\rm s}}\right]\!.
\end{equation}
Given the dark matter distribution, the cluster gas distribution
law can be found (refined). In particular, in several papers it
was proposed to modify the $\beta$ density distribution by
including a central cusp and an additional wider component (see,
e.g., Vikhlinin et al. 2006; Arnaud et al. 2010). Because of the
larger number of parameters, the modified distribution acquired
greater freedom in changing the shape and successfully fitted
the observed brightness profiles of clusters. A different
approach was realized by Shi and Komatsu (2014) and Shi et
al. (2016), who used the hydrostatic equilibrium equation for
the gas in the gravitational field of dark matter to find the
cluster gas distribution. However, they obtained the gas density
(pressure) profiles in an explicit form only for specific
temperature distributions. As will be shown below, this approach
itself can also be extended to the clusters with a constant
temperature investigated in this paper.

Neglecting the gas self-gravity, the hydrostatic equilibrium
equation has a simple form:
\begin{equation}\label{eq:nfw.hse}
 M_{\rm D}(<R)=-\frac{2kT_{\rm e}}{Gm_{\rm
     p}}\frac{d\,\ln(N_{e})}{d\,\ln(R)} R,
\end{equation}
where $G$ is the gravitational constant and $m_{\rm p}$ is the
proton mass. Substituting (\ref{eq:nfw.md}) into this equation
and integrating the resulting equation, we find the
cluster gas density distribution
\begin{equation}\label{eq:nfw.prof}
  N_{\rm e} (R)=N_{\rm s}\, e^{-\epsilon} \left(1+\frac{R}{R_{\rm
      s}}\right)^{\epsilon\,R_{\rm s}/R}.
\end{equation}
Here, $N_{\rm s}$ is the central gas electron density;
generally speaking, it is not directly related to the central
dark matter density $\rho_{\rm s}$.  The shape of the
distribution is defined by only one dimensionless parameter
\begin{equation}\label{eq:nfw.epsilon}
  \epsilon=2\pi\rho_{\rm s}R_{\rm s}^2 (Gm_{\rm p}/kT_{\rm e}).
\end{equation}
Using the characteristic values of the cluster radius $R_{500}$,
mass $M_{500}=500\, \rho_{\rm cr}(z)(4/3) \pi R_{500}^3,$ and
the concentration parameter $c$ proposed by Navarro et
al. (1997) to describe the cluster properties:
\begin{equation}\nonumber
  \begin{array}{c@{\,}c@{\,}l}
  R_{500}&=&c\,R_{\rm s}\\ [2mm]
  M_{500}&=&4\pi\rho_{\rm s}R_{\rm s}^3 E(c)^{-1},\ \mbox{\rm where}\\ [2mm]
  E(c)&=&\left[\ln(1+c)-c/(1+c)\right]^{-1},\\
  \end{array}
\end{equation}
and $\rho_{\rm cr}(z)=3/(8\pi G) H(z)^2$ is the mean (critical)
density of the Universe at cluster redshift $z$, we will
transform (\ref{eq:nfw.epsilon}) to
\begin{equation}\label{eq:nfw.epsilon2}
  \epsilon=\frac{GM_{500} m_{\rm p}}{2kT_{\rm  e} R_{\rm s}}\,E(c)\simeq
\end{equation}
\begin{equation}\nonumber
%  \begin{array}{c@{\,}c@{\,}l}
  9.1 c E(c)
  \left(\frac{M_{500}}{10^{15}\ M_{\odot}}\right)
  \left(\frac{R_{500}}{500\ \mbox{\rm kpc}}\right)^{-1}
  \left(\frac{kT_{\rm e}}{5\ \mbox{keV}}\right)^{-1}\!.
%  \end{array}
\end{equation}

Figure\,\ref{fig:nfw.prof} shows how the equilibrium gas density
distributions in a cluster with the NFW dark matter profile
calculated from Eqs.\,(\ref{eq:nfw.prof}) and
(\ref{eq:nfw.epsilon}) depend on the cluster mass $M_{500}$ and
gas temperature $kT_{\rm e}$. The concentration parameter $c$
was taken to be $2.5$. For the cluster with $M_{500}=5\times
10^{14}\ M_{\odot}$ the central gas density $N_{\rm s}$ was
determined by assuming the mass of the gas contained within the
radius $R_{500}$ to be $M_{\rm g}(R<R_{500})\simeq 0.1\ M_{500}$
(Kravtsov et al.  2006). The central gas density in the
remaining clusters was fixed at this value for the convenience
of tracing the change in profile shape (obviously, the condition
$M_{\rm g}\simeq 0.1\ M_{500}$ no longer must be fulfilled in
this case). The gas distribution depends on $kT_{\rm e}$ as
strongly as it does on $M_{500}$ (compare the distribution for
the cluster with $M_{500}=5\times 10^{14}\ M_{\odot}$ and a gas
temperature $kT_{\rm e}=5$ keV indicated by the solid blue line
and the distribution for the same cluster, but with $kT_{\rm
  e}=6$ keV, indicated by the dashed black line). This would be
expected, because these quantities enter into
(\ref{eq:nfw.epsilon2}) in the combination $M_{500}/kT_{\rm e}$.
%---------------------------------------------------------------------------------------------
\begin{figure}[h]
%\epsfxsize=0.99\linewidth
%\epsffile{szxray_z4.ps}
\centerline{\includegraphics[width=0.99\linewidth]{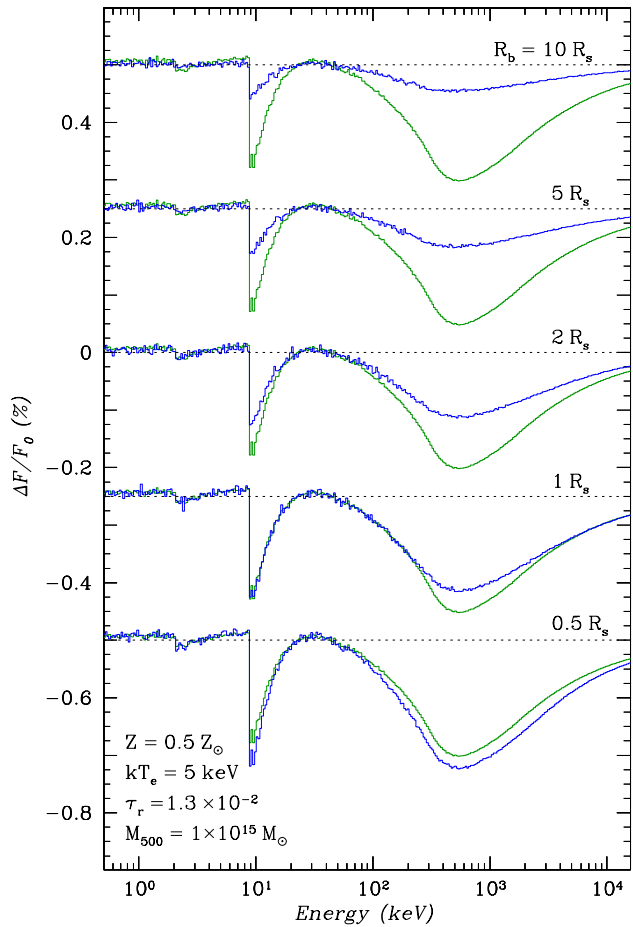}}
%\fbox{\rule{0cm}{6cm}\rule{0.97\linewidth}{0cm}}
\caption{\rm The relative background distortions (thick blue
  lines) arising in a cluster where the dark matter is
  distributed according to the NFW law, while the gas ---
  according to Eq.\,(\protect\ref{eq:nfw.prof}). The cluster
  mass is $M_{500}=1\times10^{15}\ M_{\odot}$ (the core radius
  is $R_{\rm c}\simeq 300$ kpc), the radial optical depth of the
  gas is $\tau_{\rm r}\simeq 1.3\times10^{-2}$. The cases with
  various break radii $R_{\rm b}$ are considered. The background
  distortions in a cluster with a uniform density distribution
  (thin green lines) with a radial optical depth $\tau_{\rm
    c}=6\times10^{-3}$ are shown for comparison. The gas
  temperature and metallicity in both cases are the same,
  $kT_{\rm e}=5$ keV and $Z=0.5\,Z_{\odot}$.\label{fig:nfw}}
\vspace{-4mm} 
\end{figure}
%-------------------------------------------------------------------------------------------

The thick (green) line in the figure indicates the $\beta$ gas
distribution with $\beta=2/3$ and $r_{\rm c}=0.5\ R_{\rm
  s}$. The characteristic values of $R_{\rm c}$ and $R_{500}$
are marked in the figure by the vertical dotted (red) lines. The
normalization of the gas density was chosen in such a way that
this distribution roughly coincided with the gas distribution in
the NFW model near $R\sim R_{\rm s}.$ The $\beta$ distribution
is seen to have a much gentler (flatter) profile in the central
($R\la R_{\rm s}$) part of the cluster than the NFW profile. For
comparison, the dotted (black) curve in the figure indicates the
NFW dark matter distribution proper, its normalization was set
equal to 10\% of the density $\rho_{\rm D}(R)/m_{\rm p}$ for
$M_{500}=5\times10^{14}\ M_{\odot}$\footnote{This is how the
  cluster gas would be distributed if the shape of its radial
  density profile closely coincided with the shape of the dark
  matter profile.}. The radial Thomson optical depth for such a
cluster computed by integrating the distribution
(\ref{eq:nfw.prof}) is $\tau_{\rm r}\simeq 3.3\times10^{-3},$
which accounts for $\sim 2/3$ of $\tau_{\rm c} = \sigma_{\rm T}
N_{\rm s}R_{\rm c},$ and is even smaller for more massive
clusters. For the $\beta$ distribution with a flat top the
radial optical depth is $0.5\,\pi\,\tau_{\rm c}$ (see
Eq. (\ref{eq:kingtau})). This difference is related to the more
rapid drop in gas density with radius in the central part of the
NFW model cluster.

The results of our computations of the cosmic background
distortions for a cluster with the NFW dark matter profile are
indicated in Fig.\,\ref{fig:nfw} by the thick (blue) lines. Here
we consider a more massive cluster with $M_{500}=1\times
10^{15}\ M_{\odot}$, a radial optical depth $\tau_{\rm r}=
1.3\times 10^{-2}$ ($\tau_{\rm c}=3.8\times 10^{-2}$), $R_{\rm
  s}=600$ kpc ($R_{\rm c}=300$ kpc), a gas temperature $kT_{\rm
  e}=5$ keV, and metallicity $Z=0.5\, Z_{\odot}$. For
comparison, the thin (green) lines indicate the results of our
computations of the background distortions for a cluster with a
uniform gas density with a radial optical depth $\tau_{\rm
  c}=\sigma_{\rm T} N_{\rm e} R_{\rm c}=6\times10^{-3}$ and the
same temperature $kT_{\rm e}=5$ keV. We considered the cases
with various density break radii $R_{\rm
  b}=0.5,\ 1,\ 2,\ 5,\ \mbox{and}\ 10\ R_{\rm s}.$ Although the
real Thomson optical depth of the gas in the NFW cluster exceeds
that in the homogeneous model by a factor of $\sim2$, the
distortions arising in the NFW model due to the strong gas
concentration to the center are comparable to the distortions in
a cluster with a uniform density (for the computations with
$R_{\rm b}\leq\,1R_{\rm s}$). For larger $R_{\rm b}$ the
distortion amplitude drops in the same way as in the case of a
cluster with a $\beta$ density distribution
(Fig.\,\ref{fig:king}). Obviously, the distortion depth is
determined by the optical depth averaged over the visible part
of the cluster $<\!\tau_{\rm T}\!>$. Just as in the case of a
$\beta$ distribution, it is clear that the efficiency of
background distortion observations by a telescope with an
aperture (angular resolution) with a radius more than
$\sim1-2\ R_{\rm c}$ (at given $z$) should drop rapidly.
\end{appendix}

%****************************************************************

\vspace{7mm}

~\hfill {\it Translated by V. Astakhov}
\end{document}